\documentclass[10pt,twocolumn]{article}
\usepackage[utf8]{inputenc}
\usepackage[T1]{fontenc}
\usepackage[
  letterpaper,
  top=0.85in,
  bottom=0.9in,
  left=0.75in,
  right=0.75in
]{geometry}
\usepackage{amsmath,amssymb,amsthm}
\usepackage{graphicx}
\usepackage{microtype}
\usepackage{array}
\usepackage{booktabs}
\usepackage{threeparttable}
\usepackage{tabularx}
\newcolumntype{Y}{>{\raggedright\arraybackslash}X}
\usepackage{seqsplit}
\usepackage{float}
\usepackage{adjustbox}
\usepackage{pdflscape}
\usepackage{longtable}
\usepackage{dblfloatfix}
\usepackage{balance}
\usepackage{caption}

\usepackage[round,authoryear]{natbib}
\usepackage{hyperref}
\usepackage{titlesec}
\titleformat{\section}{\normalfont\large\bfseries\raggedright}{\thesection}{0.75em}{}
\titleformat{\subsection}{\normalfont\normalsize\bfseries\raggedright}{\thesubsection}{0.75em}{}
\titleformat{\subsubsection}{\normalfont\normalsize\bfseries\raggedright}{\thesubsubsection}{0.75em}{}

\newcommand{\var}[1]{\texttt{\seqsplit{#1}}}

\newif\ifchapfivetrack
\chapfivetrackfalse
\usepackage[normalem]{ulem}
\usepackage{xcolor}

\ifchapfivetrack

\fi

\title{Revisiting the Rules-versus-Discretion Debate: Forecasting Evidence from the Bank of England, 1870--1913}
\author{Xi Chen\textsuperscript{a,\,*}\\
\textsuperscript{a}Independent researcher, Gainesville, Florida, USA\\
\textsuperscript{*}Corresponding author. Email: \texttt{100172638xi@gmail.com}.}
\date{\today}
\begin{document}

\maketitle

\begin{abstract}
Under the classical gold standard, historians have long debated what guided the Bank of England's adjustments to Bank Rate: whether policy followed a mechanical gold-standard rule or discretion when City or exchange conditions required it, and whether domestic money-market stability or external convertibility carried greater weight. This paper revisits those questions with a London-specific monthly panel, 1870--1913, jointly testing ten domestic and external indicators the Court watched when setting the published minimum rate.

The findings fit neither a single frozen channel nor unstructured ad hoc policy. Instead, the Bank appears to have drawn repeatedly on a multivariate information set---trade settlement, gold movements, City asset prices, bill-market conditions, and episodic convertibility pressure---with state-dependent weights across subperiods. In the confirmatory specification, exports, gold flow, and industrial share prices carry the clearest joint forecasting content for changes in Bank Rate. Gold flow forecasts rate changes even when reserve stock does not, separating bullion settlement from balance-sheet position. Bill-market and convertibility pressures also carry incremental forecasting content in the multivariate specification, though their prominence depends on how domestic and external indicators are modeled together. The overall pattern is most consistent with systematic discretion under the classical gold standard.
\end{abstract}

\section{Introduction}
In the classical gold standard era, 1870--1913, London was the world's leading financial center. Its financial policies shaped the global monetary system. At the center of this system was the Bank of England. As \citet{Keynes1930} described, the Bank was widely regarded as the ``conductor of the international orchestra,'' influencing interest rates, exchange rates, and capital flows worldwide. Yet despite its influence, the internal logic of Bank policy decisions remains a central debate in European economic history.

Scholars from \citet{Bloomfield1959} and \citet{Nurkse1944} through \citet{Sayers1976}, \citet{Eichengreen1992}, and \citet{Goodhart1972} have long debated the guidelines the Bank followed when setting policy: whether it mechanically obeyed gold-standard rules or exercised discretion, and whether it prioritized exchange-rate stability or domestic money-market conditions. Prior quantitative work on these questions---including monthly studies by \citet{BordoMacDonald2005} and \citet{Morys2013}---has typically tested one channel at a time or pooled countries, leaving room for a multivariate monthly analysis focused specifically on London.

This paper revisits those debates with a London-specific monthly panel that jointly tests ten domestic and external indicators the Court watched when setting the published minimum rate. The contribution is to identify which signals recurred across subperiods and how domestic City and external convertibility pressures sat on the same menu, and to evaluate sharpened mechanical and ad hoc benchmarks through rolling subsamples and lag-sensitivity tests.

\section{Literature Review}
The operation of the classical gold standard remains one of the most debated topics in European economic history. Historians disagree whether Bank policy followed mechanical gold-standard rules or systematic discretion, and whether domestic money-market or external signals dominated \citep{Bloomfield1959,Nurkse1944,Goodhart1972,Eichengreen1996,Officer1996}. The sections below review that scholarship and relate the London monthly estimates to prior quantitative work.

\subsection{Rules versus Discretion}

A central tension in the historiography is whether the Bank enforced a mechanical gold-standard rule or exercised discretion when conditions in the City or on the exchanges required it. \citet{Bloomfield1959} and \citet{Nurkse1944} describe a largely self-regulating system in which the Bank raised Bank Rate when gold flowed out and lowered it when gold returned, prioritizing exchange-rate and trade-balance stability. Their evidence rests heavily on correlations between Bank Rate and international reserves and on the logic of the price-specie-flow mechanism. \citet{Sayers1976} and \citet{Flandreau1995} counter that the Bank repeatedly placed domestic financial stability ahead of a fixed international rule book. The Baring Crisis of 1890, among other episodes documented by \citet{Sayers1976}, illustrates how policy innovations and exceptions could matter as much as reserve movements. \citet{Flandreau1995} stresses that the Bank balanced competing objectives without a mechanical rule book in such episodes. The empirical question is whether monthly data display a stable, narrow reaction function or a recurring yet re-weighted set of signals.

\subsection{Market Signals versus Exchange Rates: Triggers of Policy}

A second debate concerns which market prices the Court watched most closely when setting Bank Rate. \citet{Officer1996,Eichengreen1996} describe how, under fixed convertibility, movement of the sterling--dollar exchange rate toward the gold points called for prompt official action. \citet{Eichengreen1992} situates that mechanism in the longer history of the gold standard. \citet{Bagehot1873,Goodhart1972} instead emphasize lender-of-last-resort and money-market operations, with close attention to three-month bill and call-money rates as indicators of City stress. The two views are compatible at the level of narrative but make different predictions about which series should lead $\Delta$Bank Rate in a multivariate monthly panel. \citet{Morys2013} shows that European central banks responded to external and domestic indicators in systematically different ways. His twelve-country design does not, however, identify which London series incrementally forecast $\Delta$Bank Rate when monthly bill rates and exchange-rate movements enter the same specification.

\subsection{Domestic Stability versus International Convertibility}

The Bank's domestic and international roles need not collapse into a single priority. \citet{Eichengreen1996} stresses that convertibility constraints could lead the Bank to raise rates even when domestic political or economic conditions were unfavorable. \citet{Bagehot1873} and \citet{Goodhart1972} focus on the stability of the City: the policy rate as a tool for bill-market and banking risk in normal and crisis conditions. \citet{Flandreau1995} argues that the Bank balanced these objectives in practice, without a settled rule for which dominated in the severest panics. Monthly Granger tests \citep{Granger1969} can ask whether domestic money-market rates, the exchange rate, gold flows, and reserve stock jointly forecast official rate changes once other controls are included.

\subsection{Recent Quantitative and Institutional Research}

Since the 2000s, scholarship has returned to discount-rate policy with higher-frequency data and explicit econometric models. \citet{BordoMacDonald2005} treat the classical gold standard as a credible target zone \citep{Krugman1991} and show that core central banks, including the Bank of England, retained limited scope to set interest rates independently within the gold points even when they violated Bloomfield's ``rules of the game.'' \citet{Morys2013} compares monthly discount-rate behavior across twelve European countries and finds that core countries moved rates to defend exchange rates within the gold bands, whereas peripheral countries oriented policy toward reserve cover---a pattern that places London discount-rate decisions in a wider European system rather than in isolation.

Institutional research has likewise moved from abstract debate to the mechanics of the City. \citet{Accominotti2021} reconstruct the network of acceptors and discounters behind sterling bills of exchange using the Bank's 1906 discount ledger, showing how the Bank's rediscount operations and exposure management helped transform private bills into liquid, widely traded instruments. \citet{Ugolini2016} shows that the effectiveness of Bank Rate depended on balance-sheet tools---gold devices, Consols operations, and related practices---not on rate announcements alone. \citet{Lennard2018} uses narrative identification to ask whether monetary policy shocks had macroeconomic effects under the classical gold standard, finding large unemployment and inflation responses that earlier skepticism had understated. Together with the quantitative work above, this literature motivates asking which observable London indicators incrementally forecast $\Delta$Bank Rate when domestic money-market and external series enter jointly.

\subsection{Information Constraints and Methodological Gaps}

Finally, the empirical record has not fully caught up with the historiographical debate. \citet{Bloomfield1959} and \citet{Nurkse1944} relate Bank Rate to gold flows and reserves mainly through annual summaries and simple comovement---informative for the rules-of-the-game view, but not for asking whether one stable reaction function fits the whole classical period or whether the Bank re-weighted signals across subperiods. Monthly panel studies such as \citet{Morys2013} establish that core central banks moved discount rates with external and domestic indicators, yet their pooled European designs do not identify which London series---bill-market rates, the exchange rate, gold flows, reserve measures, and trade---incrementally forecast $\Delta$Bank Rate when entered together. The sections that follow address that predictive question on a consistent monthly London panel.

\section{Historical and Institutional Background}
\label{sec:background}

Under the classical gold standard, the Bank of England's Bank Rate was both an instrument of domestic money-market management and a signal watched across the international financial system. This section motivates the monthly indicators and the historical channels they represent.

\subsection{The Bank of England and the London Money Market}

London was the principal international financial center of the era. A dense network of bill finance, merchant banking, and short-term lending centered in the City meant that conditions in the London money market could change quickly when credit tightened or confidence weakened. The Bank of England sat at the center of this system. It was not merely an issuer of notes. Through its discount window and its influence over the terms at which high-quality bills could be turned into liquidity, it could affect the marginal cost of short-term credit \citep{Goodhart1972,Sayers1976}.

Bank Rate---the official minimum rate at which the Bank discounted first-class bills---was the most visible policy tool. Changes in Bank Rate altered the opportunity cost of obtaining liquidity at the Bank and, through arbitrage and market practice, were linked to other short-term rates in the City. Historians of operational central banking therefore treat rates on short-dated bills and money-market liquidity as proximate indicators of stress \citep{Goodhart1972,Accominotti2021}. In the empirical analysis below, the London open-market three-month discount rate and the prime six-month bill rate proxy City money-market conditions; month-to-month changes in the published Bank Rate minimum serve as the response variable, including months with no change; see Table~\ref{tab:hypotheses}.

\subsection{How Bank Rate Was Set in Practice}

Bank Rate was set by the Court of Directors, typically on Thursdays, after the Governor and senior staff reported on reserves, the foreign exchanges, and conditions in the discount market \citep{Sayers1976,Goodhart1972}. The Court did not follow a published formula. Directors weighed gold flows, sterling's position against the dollar and other currencies, and the tone of the bill market before announcing a new minimum rate at which the Bank would lend on eligible paper. In calm years Bank Rate could remain unchanged for months; in crises it might move several times within a quarter, as in 1890 \citep{Flandreau1995} and 1907 \citep{Ugolini2016}.

\citet{Sayers1976,Goodhart1972} report that contemporaries and later historians treated the published rate as a benchmark that pulled bill rates and call money with it, but not instantly or perfectly. \citet{Bagehot1873} described how panic could force the Bank to lend freely even while keeping a high penalty rate. \citet{Accominotti2021} shows how acceptance houses and discounters stood between the Bank and the wider market, shaping how quickly stress in private bills translated into pressure on official rate policy. The monthly data used here capture month-to-month changes in the published minimum, including months with no change, and permit comparison of lagged movements in market rates, the exchange rate, reserves, and macro indicators with those adjustments.

\subsection{The Gold Standard and External Constraints}

The classical gold standard tied sterling to gold at a fixed parity and made convertibility a central constraint on policy. When pressure on sterling or on the Bank's gold reserves intensified, market participants and other central banks interpreted Bank Rate changes as signals about the Bank's willingness and ability to defend the standard \citep{Eichengreen1992,Eichengreen1996}. The structure of raising Bank Rate when gold flowed out and lowering it when gold flowed in describes one automatic mapping from external imbalance to policy \citep{Bloomfield1959,Nurkse1944}. For present purposes, the important point is that external pressure is plausibly reflected in the exchange rate and in measures of reserve adequacy.

When the sterling--dollar exchange rate moved toward the edges of the gold points---the band around mint parity within which it was not profitable to ship gold \citep{Eichengreen1992,Officer1996}---arbitrage and capital flows could accelerate, affecting reserves and the urgency of policy responses. The empirical analysis includes the sterling--dollar exchange rate and net gold flows as indicators of the external-convertibility channel, together with Bank of England (BoE) Banking Department reserve stock and monthly imports and exports. Under the gold standard, rising imports increased demand for sterling bills to finance trade and could tighten external balance before reserve or exchange-rate indicators fully reflected the pressure \citep{Bloomfield1959,Accominotti2021}. Industrial share prices, gilt prices, and wheat prices capture broader financial and real-economy conditions; see Table~\ref{tab:hypotheses}.

\subsection{Transmission: How Bank Rate Reached Markets and Reserves}

Three channels help connect Bank Rate to the predictor variables used in this study.

Domestic liquidity. By raising or lowering the discount rate, the Bank influenced the terms on which the City obtained liquidity. Tighter Bank Rate settings tended to coincide with higher bill and call rates as institutions adjusted across the short end of the market. Historians treat the three-month rate on eligible bills as the operational benchmark the Court watched most closely \citep{Goodhart1972,Sayers1976}. Lagged changes in bill-market rates are therefore natural candidates to enter a multivariate reaction function for $\Delta$Bank Rate.

International arbitrage and trade finance. Interest-rate differentials between London and New York and other centers affected capital flows and gold movements under the gold standard. Rising imports could likewise tighten external balance by increasing the commercial demand for bills and sterling, sometimes preceding visible reserve loss or exchange-rate movement. If external convertibility mattered, month-to-month changes in the sterling--dollar rate, net gold flows, and reserve stock are natural candidates to enter jointly when pressure on sterling and gold movements intensified together. Merchandise trade also follows pronounced calendar seasonality at monthly frequency, so imports and exports may embed recurring calendar rhythms alongside payment-pressure signals.

Real-economy controls. The Court's deliberations centered on the discount market and the exchanges rather than industrial output or wholesale prices \citep{Sayers1976,Goodhart1972}. Month-to-month changes in industrial share prices, gilt prices, and wheat prices enter as financial and commodity proxies for broader conditions; they may matter episodically but are not expected to dominate bill and external indicators in a multivariate reaction function.

Crisis and lender-of-last-resort conditions. \citet{Bagehot1873} famously argued that in a panic the central bank should lend freely against good collateral at a penalty rate. The Bank did not always follow that prescription literally, but the doctrine highlights why domestic market rates could spike in crises even when the international picture was also strained. Domestic and external indicators often moved together in such episodes, which motivates estimating a multivariate model rather than relying on pairwise correlations alone.

\subsection{Crisis Episodes, 1870--1913}

The sample spans the full classical gold standard period and includes three widely discussed stress episodes \citep{Flandreau1995,Ugolini2016,ObstfeldTaylor2004}. These are pre-specified reference periods for interpreting whether predictive relationships strengthen or shift in rolling windows that overlap them.

The Panic of 1873. Triggered by the collapse of Jay Cooke \& Co.\ in the United States and a crash on the Vienna Stock Exchange, the panic ended a global railroad boom and contributed to a prolonged trade depression in Britain \citep{ObstfeldTaylor2004}. It marks an early subperiod in the sample in which both domestic and international conditions deteriorated.

The Baring Crisis of 1890. When Baring Brothers, one of London's leading merchant banks, faced insolvency from exposure to defaulted Argentine sovereign debt, the Bank of England organized a private bailout of roughly \pounds 17 million. The crisis was centered in the City and is often cited as an example of discretionary crisis management rather than a mechanical response to gold outflows alone \citep{Flandreau1995}.

The Panic of 1907. A liquidity freeze in the United States led American banks to draw gold from London. The Bank of England raised Bank Rate to 7\%, illustrating how external pressure and domestic tight money could coincide at the end of the sample \citep{Ugolini2016}.

\subsection{Institutional Channels and Indicators}

The institutional background above maps into the monthly indicators used in the analysis. Table~\ref{tab:hypotheses} links each narrative from \S\S3.1--3.4 to its measured series.

\begin{table}[!t]
\centering
\caption{Historical narratives and primary indicators}
\label{tab:hypotheses}
\footnotesize
\begin{tabular}{@{}>{\raggedright\arraybackslash}p{0.36\columnwidth}>{\raggedright\arraybackslash}p{0.60\columnwidth}@{}}
\toprule
Narrative & Primary indicators \\
\midrule
Domestic money-market \citep{Bagehot1873,Goodhart1972} &
\var{Market\_Rate\_3Month\_diff}, \var{Market\_Rate\_6Month\_diff} \\
\midrule
External convertibility \citep{Bloomfield1959,Eichengreen1992} &
\var{Exchange\_Rate\_USD\_diff}, \var{Gold\_Flow\_diff}, \var{BoE\_Reserve\_Notes\_Coin\_diff} \\
\midrule
Trade and external payments \citep{Bloomfield1959} &
\var{Imports\_diff}, \var{Exports\_diff} \\
\midrule
Financial and commodity proxies &
\var{Industrial\_Share\_Price\_Index\_diff}, \var{Gilt\_Price\_Index\_diff}, \var{Wheat\_Price\_diff} \\
\midrule
Systematic discretion, this study &
Mechanical vs.\ ad hoc extremes; all predictors; rolling subsamples \\
\bottomrule
\end{tabular}
\end{table}

The following section describes the data sources, transformations, and Granger-causality framework used with these indicators.

\section{Methodology}

This section describes how the indicators in Section~\ref{sec:background} are mapped into data, transformations, and time-series models. The empirical goal is to identify which observable indicators contain incremental predictive information for month-to-month changes in Bank Rate, and how stable those relationships are across subperiods \citep{Granger1969,Hamilton1994}.

\subsection{Data and Variables}
\label{sec:data}

The analysis uses monthly data from the National Bureau of Economic Research (NBER) Macrohistory Database and complementary archival sources, following the compilations in \citet{Mitchell1988} and \citet{CapieWebber1985} and the cliometric tradition in recent gold-standard central-banking work \citep{BordoMacDonald2005,Morys2013,Lennard2018}. The panel runs from January 1870 to December 1913, spans 528 months, and comprises eleven monthly series: ten predictors and Bank Rate. Appendix~\ref{app:data_sources} records series identifiers, units, and non-NBER sources, including Federal Reserve Economic Data (FRED), Bank of England Millennium data, and the BoE weekly balance sheet. The outcome is month-to-month changes in the published Bank Rate minimum. Predictors comprise London three- and six-month bill rates, the sterling--dollar exchange rate, net gold flows, Banking Department reserve stock, imports and exports, and proxies for wheat prices, industrial share prices, and gilt prices. Rates are in percent per annum unless noted. All series enter in first differences. Unit-root tests that motivate that transformation are reported in the next subsection. Granger regressions with six lags employ 521 observations after differencing and lag truncation.

Three measurement choices matter for interpretation. First, Bank Rate is the published minimum discount rate, not a weekly high or a within-month average, so month-to-month changes summarize within-month Court decisions rather than a within-month average quote. Second, the sterling--dollar exchange rate is the monthly cable quote in dollars per pound; robustness checks replace this proxy with gold-point pressure and parity deviation constructed from the same underlying quote. Third, net gold bullion and specie trade and Banking Department reserve stock are distinct external indicators: bullion settlement through the trade account versus balance-sheet reserve position. Appendix~\ref{app:data_sources} records the corresponding source identifiers; Appendix~\ref{app:robustness_extra} reports a reserve-to-liabilities ratio from 1876 as a further balance-sheet check.

\subsection{Stationarity Testing and Data Transformation}
\label{sec:stationarity}

The main Granger regressions use first differences of all eleven series, reflecting mixed time-series properties in the full monthly panel.

Augmented Dickey--Fuller (ADF) tests \citep{DickeyFuller1979} ask whether each series is stationary in levels---that is, whether it fluctuates around a stable mean rather than drifting persistently over decades. On the full predictor set, ADF rejects stationarity at the 5\% level for seven series, including imports, exports, gilt and industrial share price indices, wheat prices, the exchange rate, and BoE reserve stock. Bank Rate, the London bill rates, and gold flow are stationary in levels, consistent with policy and market bounds under the gold standard.

Regressions that mix persistent macro variables in levels with mean-reverting rates can nonetheless produce misleading significance when series share common trends \citep{GrangerNewbold1974}. In our data, initial Granger tests on levels flagged the industrial share price index and gilt price index at the 1\% level. Those results vanish after differencing.

We therefore apply first differences ($\Delta X_t = X_t - X_{t-1}$) to all variables in the primary analysis, including rates that are already stationary in levels. Every predictor then enters the ten-predictor multiple linear regression (MLR) as month-to-month changes, which aligns the response variable with discrete policy moves in $\Delta$Bank Rate and avoids mixing levels with differences in one equation. ADF tests on every differenced series reject a unit root at the 5\% level. The headline reaction-function question is therefore addressed throughout in month-to-month changes.

\subsection{Econometric Approach: Granger Causality}
\label{sec:econometric}
\label{sec:inferential_hierarchy}

The empirical design follows Table~\ref{tab:hypotheses} in four layers: exploratory screening, confirmatory inference, robustness checks, and multivariate complements. Robustness repeats the confirmatory MLR under alternative lag depths, seasonal settings, bill-rate combinations, external-pressure measures, multivariate systems, subsamples, and discrete-outcome models (Sections~\ref{sec:diagnostics}, \ref{sec:var}, \ref{sec:spec_checks}, and~\ref{sec:stability_oos}, with supplementary tables in Appendix~\ref{app:robustness_extra}). The subsections below set out the Granger framework and each specification.

We use Granger-causality tests \citep{Granger1969,Sims1980} within a multiple linear regression framework to ask whether lagged predictors improve forecasts of $\Delta$Bank Rate beyond Bank Rate's own past and the other controls in the equation.

\subsubsection{What ``Granger Cause'' Means Here}

Throughout this paper, ``$X$ Granger-causes Bank Rate'' means lagged changes in $X$ improve forecasts of $\Delta$Bank Rate beyond Bank Rate's own lags and the other variables in the model. Several predictors moved with Bank Rate in the same macro environment; the MLR asks which series retain marginal forecasting content conditional on the other indicators in Table~\ref{tab:hypotheses}, and the compact differenced VAR in Section~\ref{sec:methods_multivariate} models short-run feedback among Bank Rate, bill-market rates, the exchange rate, and gold flow.

\subsubsection{Model Specifications}

Granger causality compares two linear regression specifications.

The restricted model uses lagged changes in Bank Rate alone to forecast the current change ($\Delta Y_t$):
\begin{equation}
\Delta Y_t = \alpha + \sum_{i=1}^{L} \beta_i \Delta Y_{t-i} + \varepsilon_t
\end{equation}

The unrestricted model adds lagged changes in a predictor to lagged changes in Bank Rate:
\begin{equation}
\Delta Y_t = \alpha + \sum_{i=1}^{L} \beta_i \Delta Y_{t-i} + \sum_{i=1}^{L} \gamma_i \Delta X_{t-i} + \varepsilon_t
\end{equation}

We incorporate lags of one to six months. At maximum lag six, the regression employs 521 observations after differencing and lag truncation.

\subsubsection{Multiple Linear Regression and Confounding Variables}

We extend the bivariate setup to an MLR that includes all predictors at once. When the coefficient on $\Delta X_n$ is significant even with lags of other variables $\Delta Z_n$ in the model, that predictor still helps forecast $\Delta$Bank Rate beyond shared market movements.

The null hypothesis states that the regression coefficients on all lags of a predictor are zero. We test that restriction with an $F$-test in the bivariate setup and with joint block tests in the confirmatory multivariate specification.

\subsubsection{Exploratory screening and temporal validation}

For each maximum lag $L \in \{1,\ldots,6\}$, we estimate a separate ten-predictor MLR and record, for each predictor, the smallest $p$-value across lags $1$--$L$, summarized in a composite heatmap. We then apply Bonferroni and false discovery rate (FDR) adjustments \citep{BenjaminiHochberg1995} to the ten minimum $p$-values at each lag depth, repeat the same logic in rolling subsamples, and turn to confirmatory block tests in the next subsection.

\noindent
Bonferroni divides the significance level by the number of tests, controlling the family-wise error rate. FDR controls the expected proportion of false discoveries and is less conservative than Bonferroni.

To assess whether significant terms are stable across the gold-standard era or confined to particular episodes, we implement rolling-window analysis on the January 1870--December 1913 panel: 37 separate regressions on seven-year, 84-month spans stepped forward one year at a time. Three stress episodes noted above---the Panic of 1873, the Baring Crisis of 1890, and the Panic of 1907---serve as reference periods when interpreting windows that overlap them. Each subsample uses $L=3$ so the number of parameters remains manageable within shorter windows. This design helps identify temporal stability and potential regime changes.

As a further check on economic plausibility, we generate dual-axis time series plots for each predictor that registers at least one Bonferroni- or FDR-significant term--lag pair in the exploratory MLR, lagged at one through six months, against $\Delta$Bank Rate. These figures provide intuitive validation of temporal precedence alongside the regression evidence in Appendix~\ref{app:timeseries}.

\subsubsection{Confirmatory lag length and formal inference}
\label{sec:lag_strategy}

Once the exploratory screen has highlighted candidate predictors and lag patterns, we hold the econometric specification fixed and conduct formal inference on that single design. Confirmatory inference uses a ten-predictor MLR at maximum lag $L=6$, with eleven monthly seasonal dummies from February through December and January omitted, and Newey--West heteroskedasticity- and autocorrelation-consistent (HAC) standard errors \citep{NeweyWest1987}. We set the HAC bandwidth to six lags, matching the maximum lag depth in the regression. For each predictor we test whether all of its lags, taken together, improve forecasts of $\Delta$Bank Rate beyond the other variables in the model. A no-seasonal specification serves as a calendar-structure robustness check.

The Court typically adjusted Bank Rate on a weekly rhythm, while this study uses end-of-month observations for the multivariate panel. Monthly $\Delta$Bank Rate therefore aggregates within-month decisions.

We use $L=6$ as the primary confirmatory lag length because it matches the longest order considered in the exploratory search. Information criteria for the seasonal MLR point in different directions: the Bayesian information criterion (BIC) and Hannan--Quinn information criterion (HQIC) select $L=1$ and the Akaike information criterion (AIC) selects $L=4$ (Appendix~\ref{app:confirmatory}, Table~\ref{tab:ch2_lag_ic}). We repeat joint block tests at $L=1$, $4$, and $6$ and extend the same lag depths to pseudo-real-time forecast evaluation.

\subsubsection{Compact multivariate system}
\label{sec:methods_multivariate}

The MLR treats Bank Rate as the sole outcome and conditions on ten predictors at once, which is appropriate for asking which channels retain marginal predictive content in a rich reaction function. We also estimate a parsimonious four-variable system with simultaneous feedback among Bank Rate, the three-month London market rate, the US dollar/British pound sterling (USD/GBP) exchange rate, and net gold bullion and specie trade. These series map directly onto the domestic money-market and external-convertibility rows in Table~\ref{tab:hypotheses}. The three-month rate serves as the domestic money-market proxy in the four-variable system. The six-month bill rate is highly correlated with it and remains in the ten-predictor specification rather than expanding a vector autoregression (VAR). BoE reserve stock, imports, exports, and financial proxies enter only the MLR, where their incremental content can be assessed without inflating a simultaneous system that would be poorly identified at monthly frequency with many lags.

We use two four-variable systems, each specified according to its integration properties (Appendix~\ref{app:var_setup}). In the policy block---Bank Rate, three-month market rate, USD/GBP, and gold flow---Bank Rate, the three-month rate, and gold flow are stationary in levels while the exchange rate is nonstationary in levels. We therefore estimate a differenced VAR with BIC lag selection and apply Toda--Yamamoto Granger tests \citep{TodaYamamoto1995} on levels. The external-adjustment block---USD/GBP, imports, exports, and BoE reserve stock---is nonstationary in levels throughout and is analyzed with Johansen trace tests and a vector error correction model (VECM) \citep{Johansen1991,EngleGranger1987} to test whether trade, sterling pressure, and reserve stock share a long-run external-balance relation consistent with Table~\ref{tab:hypotheses}, with short-run VECM Granger tests on reserve-stock dynamics. Net gold trade enters the policy block rather than the VECM because it is mean-reverting in levels.

\section{Results}
\label{sec:results}

This section evaluates the indicators in Table~\ref{tab:hypotheses}. We ask which monthly series forecast $\Delta$Bank Rate---domestic money-market stress versus external convertibility pressure---and whether those leads are stable across subperiods. We begin with exploratory and rolling screens, turn to confirmatory block tests and magnitudes in the canonical MLR, examine the domestic--external core in a compact multivariate system, and then report specification checks on measurement and discrete rate moves, subperiod stability, and pseudo-real-time forecasting.

\subsection{Full-Sample Analysis}

\subsubsection{Overall Summary}

\begin{figure*}[!t]
\centering
\adjustbox{width=\textwidth,max height=0.36\textheight,center}{\includegraphics{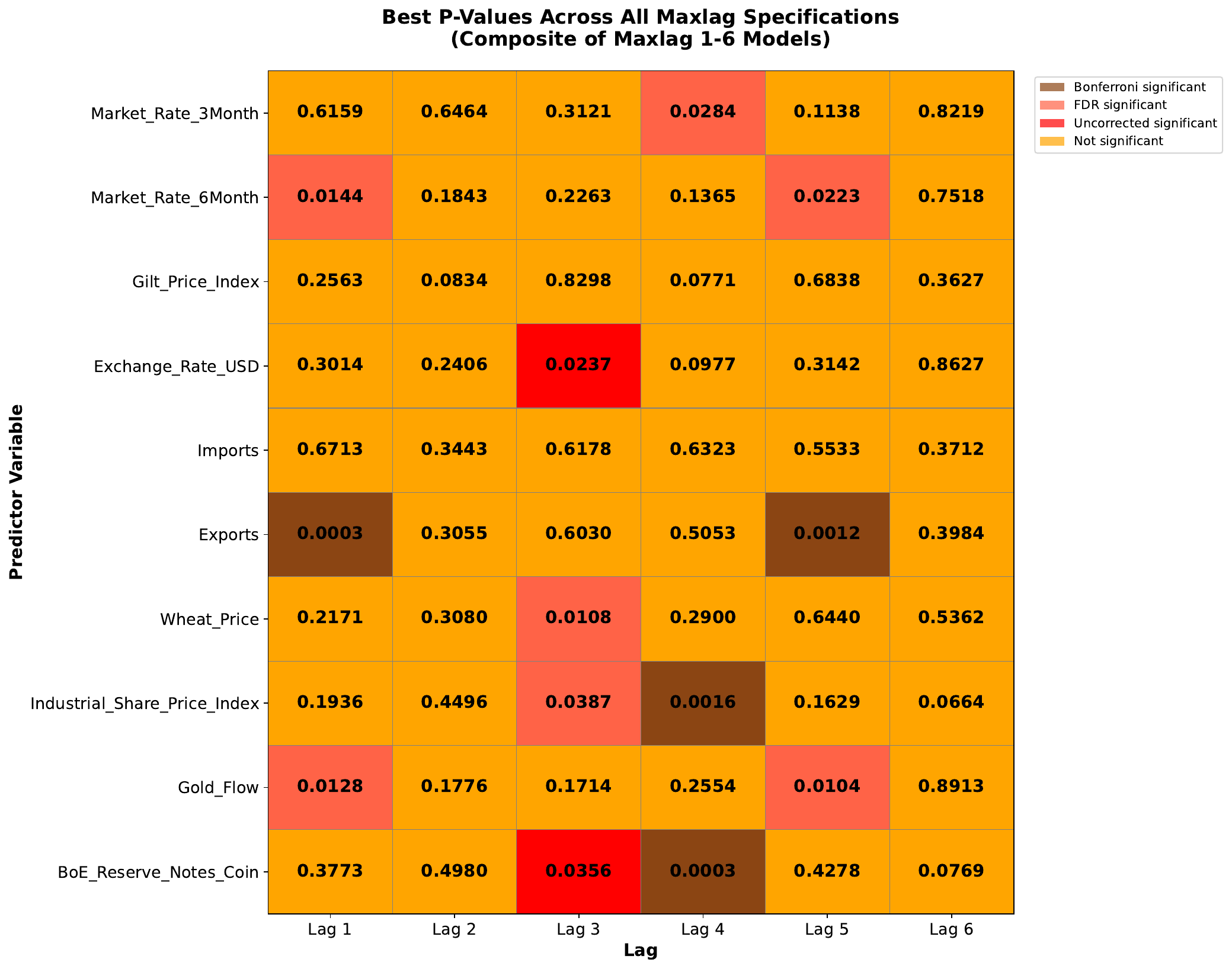}}
\caption{Exploratory screen: minimum lag-specific Granger $p$-value by predictor and maximum lag depth. Brown: Bonferroni; coral: FDR; red: uncorrected 5\%; orange: not significant.}
\label{fig:composite_heatmap}
\end{figure*}

Figure~\ref{fig:composite_heatmap} summarizes the exploratory screen described in Section~\ref{sec:econometric}: for each maximum lag depth $L=1,\ldots,6$, we record the smallest lag-specific $p$-value for each of ten predictors and apply Bonferroni and FDR adjustments across those ten summaries. Exports dominate Bonferroni passes across lag depths; gold flow and the six-month bill rate register mainly under FDR; industrial share prices and BoE reserve stock appear under both corrections; see Appendix~\ref{app:exploratory} and Table~\ref{tab:exploratory_comprehensive}. The six-month bill rate registers more exploratory hits than the three-month rate; the latter appears mainly at lags four--five. Section~\ref{sec:diagnostics} reports confirmatory HAC block tests on the full lag set of each predictor.

\subsection{Rolling Window Analysis}

To assess temporal stability, the analysis was repeated across 37 rolling windows of seven years each, stepped forward one year over January 1870--December 1913.

\subsubsection{Overall Statistics}

Across 37 windows and ten predictors, a minority of screening outcomes pass under strict correction. Table~\ref{tab:rolling_window_freq} reports how often each predictor passes at the uncorrected, FDR, and Bonferroni tiers. Appendix~\ref{app:rolling}, Figure~\ref{fig:rolling_raw}, displays the full uncorrected rolling-window heatmap.

\subsubsection{Most Consistent Predictors Across Windows}

BoE reserve stock and gilt prices appear in the largest share of uncorrected windows, 13 of 37 and 11 of 37 respectively. Exports, the six-month bill rate, gold flow, imports, and industrial share prices register in 7--8 windows uncorrected but rarely under Bonferroni. The three-month rate appears in only 5 of 37 windows uncorrected and in none after FDR or Bonferroni correction. No predictor is significant in every window at every threshold, consistent with state-dependent weights on the signals in Table~\ref{tab:hypotheses}.

\begin{table}[!t]
\centering
\footnotesize
\begin{threeparttable}
\caption{Rolling-window screen: share of 37 seven-year windows with significant predictor ($L=3$)}
\label{tab:rolling_window_freq}
\setlength{\tabcolsep}{4pt}
\begin{tabular*}{\columnwidth}{@{\extracolsep{\fill}}@{}lrrr@{}}
\toprule
Predictor & Uncorr. & FDR & Bonf. \\
\midrule
3m market rate & 5/37 & 0/37 & 0/37 \\
6m market rate & 8/37 & 2/37 & 1/37 \\
USD/GBP & 6/37 & 2/37 & 1/37 \\
Gold flow & 7/37 & 1/37 & 0/37 \\
BoE reserves & 13/37 & 3/37 & 0/37 \\
Imports & 7/37 & 2/37 & 1/37 \\
Exports & 8/37 & 0/37 & 0/37 \\
Wheat price & 5/37 & 0/37 & 0/37 \\
Gilt prices & 11/37 & 4/37 & 3/37 \\
Industrial shares & 7/37 & 3/37 & 2/37 \\
\bottomrule
\end{tabular*}
\begin{tablenotes}[flushleft]
\footnotesize
\item[] 37 windows of 84 months, stepped one year (Jan 1870--Dec 1913; ten predictors). Heatmaps: Figures~\ref{fig:rolling_raw}--\ref{fig:rolling_bonf}.
\end{tablenotes}
\end{threeparttable}
\end{table}

\subsection{Confirmatory estimates and block tests}
\label{sec:diagnostics}

We report confirmatory estimates at maximum lag $L=6$, following Section~\ref{sec:lag_strategy}, with Newey--West HAC standard errors, joint block tests, and standard regression diagnostics.

\subsubsection{Canonical full-sample regression}
\label{sec:canonical}

The panel runs from January 1870 through December 1913 with ten differenced predictors, eleven monthly seasonal dummies with January omitted, and 521 observations after lag construction. Table~\ref{tab:ch2_canonical_hac} in Appendix~\ref{app:confirmatory} reports selected coefficients. Adjusted $R^2$ is about 0.32. None of the headline predictors in that table load significantly on the first lag under HAC inference at the 5\% level once all ten predictors enter. Monthly changes in the quoted \$/\pounds{} rate do not form a jointly significant block at conventional levels.

\subsubsection{Joint block tests for all predictors}

Table~\ref{tab:ch2_all_blocks} reports HAC block $p$-values for all ten predictors at six lags and 521 observations.

\begin{table}[!t]
\centering
\small
\begin{threeparttable}
\caption{HAC block Granger tests on all ten predictors ($p=6$, canonical MLR with monthly seasonal dummies)}
\label{tab:ch2_all_blocks}
\begin{tabular*}{\columnwidth}{@{\extracolsep{\fill}}lr}
\toprule
Predictor & Block $p$ (HAC) \\
\midrule
Gold flow & \textbf{\textless{}0.001} \\
Industrial share prices & \textbf{0.0016} \\
Exports & \textbf{0.0059} \\
6-month market rate & 0.0517 \\
Imports & 0.0621 \\
Wheat price & 0.1424 \\
USD/GBP exchange rate & 0.1779 \\
Gilt price index & 0.2968 \\
BoE reserve stock & 0.3314 \\
3-month market rate & 0.3339 \\
\bottomrule
\end{tabular*}
\begin{tablenotes}[flushleft]
\footnotesize
\item[] Bold entries: $p<0.05$ (two-sided).
\end{tablenotes}
\end{threeparttable}
\end{table}

\noindent
At the 5\% level, exports, industrial share prices, and gold flow are jointly significant. Imports at $p \approx 0.062$ and the six-month bill rate at $p \approx 0.052$ are marginal; the three-month bill rate, BoE reserve stock, gilt prices, wheat prices, and the cable exchange rate are not.

\subsubsection{Lag depth}
\label{sec:lag_ic}

\begin{table}[!t]
\centering
\footnotesize
\begin{threeparttable}
\caption{HAC block $p$-values by maximum lag length (canonical ten-predictor MLR with monthly seasonal dummies)}
\label{tab:lag_sensitivity}
\setlength{\tabcolsep}{3pt}
\begin{tabular*}{\columnwidth}{@{\extracolsep{\fill}}@{}lrrr@{}}
\toprule
Predictor & $p=1$ & $p=4$ & $p=6$ \\
\midrule
Gold flow & \textbf{0.0067} & \textbf{0.0052} & \textbf{\textless{}0.001} \\
Industrial share prices & 0.6807 & \textbf{0.0110} & \textbf{0.0016} \\
Exports & \textbf{0.0216} & \textbf{0.0269} & \textbf{0.0059} \\
6-month market rate & 0.0815 & 0.0519 & 0.0517 \\
Imports & 0.6441 & 0.5781 & 0.0621 \\
Wheat price & 0.1553 & 0.1109 & 0.1424 \\
USD/GBP exchange rate & 0.8803 & 0.1030 & 0.1779 \\
Gilt price index & 0.7935 & 0.2628 & 0.2968 \\
BoE reserve stock & 0.4733 & \textbf{0.0394} & 0.3314 \\
3-month market rate & 0.6096 & 0.3303 & 0.3339 \\
\bottomrule
\end{tabular*}
\begin{tablenotes}[flushleft]
\footnotesize
\item[] Bold entries: $p<0.05$ (two-sided).
\end{tablenotes}
\end{threeparttable}
\end{table}

\noindent
Information criteria favor shorter lag orders than the six-lag confirmatory specification: BIC and HQIC select one lag and AIC selects four; see Section~\ref{sec:lag_strategy} and Table~\ref{tab:ch2_lag_ic} in Appendix~\ref{app:confirmatory}. Table~\ref{tab:lag_sensitivity} reports HAC block tests for all ten predictors at maximum lags one, four, and six.

Exports and gold flow pass at all three depths shown. Industrial share prices are significant at $L=4$ and $L=6$ but not at $L=1$. BoE reserve stock is significant at $L=4$ with $p \approx 0.039$ but not at $L=1$ or $L=6$. Imports are marginal only at $L=6$ with $p \approx 0.062$; the six-month bill rate is marginal at $L=4$ and $L=6$. Exchange-rate changes, gilt prices, and wheat prices are insignificant at all three depths. Industrial share prices and reserve stock enter from four lags onward; imports and bill-market rates are the main predictors whose joint blocks strengthen only at longer depths. Section~\ref{sec:stability_oos} reports pseudo-real-time forecast metrics at the same lag depths.

\subsubsection{Residual and collinearity diagnostics}
\label{sec:residuals}

Ljung--Box tests \citep{LjungBox1978} on ordinary least squares (OLS) residuals from the canonical regression show no meaningful remaining autocorrelation; see Table~\ref{tab:ch2_ljungbox} in Appendix~\ref{app:confirmatory}. Maximum variance inflation factor (VIF) values reach about 10 for the six lags of each bill rate in Table~\ref{tab:ch2_vif}, reflecting near-collinearity between three- and six-month rates.

\subsubsection{Bill-rate collinearity}
\label{sec:bill_rate_robustness}

Table~\ref{tab:bill_rate_robustness} repeats HAC block tests when only one London bill rate enters the seasonal MLR. With both rates included, the six-month rate is marginal at $p \approx 0.052$ and the three-month rate is insignificant at $p \approx 0.33$. When each rate enters alone, both are highly significant at $p < 0.01$. The split in the baseline specification therefore reflects multicollinearity between maturities rather than absence of bill-market content.

\begin{table}[!t]
\centering
\footnotesize
\begin{threeparttable}
\caption{HAC block $p$-values: bill-rate specifications ($L=6$, other predictors unchanged)}
\label{tab:bill_rate_robustness}
\setlength{\tabcolsep}{4pt}
\begin{tabular*}{\columnwidth}{@{\extracolsep{\fill}}lr}
\toprule
Setting & Block $p$ \\
\midrule
Both: 3-month & 0.3339 \\
Both: 6-month & 0.0517 \\
3-month only & \textbf{\textless{}0.001} \\
6-month only & \textbf{\textless{}0.001} \\
\bottomrule
\end{tabular*}
\begin{tablenotes}[flushleft]
\footnotesize
\item[] Bold entries: $p<0.05$ (two-sided).
\end{tablenotes}
\end{threeparttable}
\end{table}

\subsubsection{Effect sizes for block-significant predictors}
\label{sec:effect_sizes}

Statistical significance alone does not summarize economic scale. Table~\ref{tab:effect_sizes} reports HAC lag-one coefficients, the sum of coefficients across lags one--six (a compact distributed-lag total), and that sum multiplied by one within-sample standard deviation of the predictor for the block-significant predictors in the seasonal baseline.

\begin{table}[!t]
\centering
\footnotesize
\begin{threeparttable}
\caption{HAC coefficient magnitudes for block-significant predictors ($L=6$, seasonal canonical MLR)}
\label{tab:effect_sizes}
\setlength{\tabcolsep}{3pt}
\begin{tabular*}{\columnwidth}{@{\extracolsep{\fill}}lrrr}
\toprule
Predictor & $\beta_1$ & $\sum_{j=1}^{6}\beta_j$ & 1-SD impact \\
\midrule
Gold flow & -0.000 & +0.000 & +0.056 \\
Industrial share prices & -0.010 & +0.020 & +0.033 \\
Exports & -0.032 & -0.036 & -0.068 \\
\bottomrule
\end{tabular*}
\begin{tablenotes}[flushleft]
\footnotesize
\item[] 1-SD impact is one within-sample standard deviation of the predictor times $\sum_{j=1}^{6}\beta_j$ (percentage points on $\Delta$Bank Rate).
\end{tablenotes}
\end{threeparttable}
\end{table}

\noindent
In levels, a one-standard-deviation monthly change in net gold flow of about \pounds 1.9 million in the within-sample distribution aligns with roughly six basis points on $\Delta$Bank Rate across the six-lag sum; a one-standard-deviation export movement of about \pounds 1.9 million with roughly seven basis points in the opposite direction; and a one-standard-deviation move in industrial share prices with about three basis points. These are order-of-magnitude associations from the distributed-lag totals, not structural pass-throughs. For comparison, monthly $\Delta$Bank Rate itself has a within-sample standard deviation of about 0.6 percentage points; in 1890 the largest monthly increase was about one percentage point while gold-flow swings exceeded \pounds 5 million, and in 1907 the largest monthly increase was about 2.2 percentage points during the gold outflows that preceded the 7\% November rate. The MLR therefore links recurring predictive content to scales that are modest in any single month but sit alongside the larger, episodic rate moves historians associate with the Baring and 1907 episodes \citep{Flandreau1995,Ugolini2016}.

\subsubsection{Inferential structure of the significance tests}
\label{sec:testing_family}

Section~\ref{sec:inferential_hierarchy} orders the evidence: exploratory screens generate candidates; confirmatory block tests apply to the seasonal MLR at $L=6$; robustness and multivariate sections vary measurement or system structure while keeping the core forecasting question in view.

Within the exploratory stage, we record the minimum $p$-value across lags $1$--$L$ for each predictor at each lag depth $L$, then apply Bonferroni and FDR adjustments to those ten values at that depth.

Three patterns carry into the sections below. Exports, gold flow, and industrial share prices pass confirmatory joint blocks in the seasonal baseline, with magnitudes in Table~\ref{tab:effect_sizes}. Bill-market pressure is present in both maturities but is split across them once both bill rates enter (Table~\ref{tab:bill_rate_robustness}). Raw exchange-rate changes do not pass the same block tests, although pressure outside the gold points and the London--New York spread enter in Section~\ref{sec:spec_checks}; the policy VAR and external VECM in Section~\ref{sec:var} add feedback and long-run perspectives on the same channels.

\subsection{Multivariate time-series representations}
\label{sec:var}

We complement the single-equation Granger results with the policy VAR and external VECM blocks specified in Section~\ref{sec:methods_multivariate}. The policy block asks which domestic money-market and foreign exchange (FX) companions forecast $\Delta$Bank Rate when modeled jointly with Bank Rate; the external block asks whether trade, sterling pressure, and reserve stock share a long-run convertibility relation.

\subsubsection{VAR in first differences (policy block)}

\begin{table}[!t]
\centering
\small
\begin{threeparttable}
\caption{VAR Granger causality ($p=2$ by BIC): forecasting $\Delta$Bank Rate}
\label{tab:ch2_granger_var}
\setlength{\tabcolsep}{3pt}
\begin{tabular*}{\columnwidth}{@{\extracolsep{\fill}}@{}llrr@{}}
\toprule
Block & Test & $\chi^2$ & $p$ \\
\midrule
Open-market 3m & Individual & 27.1293 & \textbf{\textless{}0.001} \\
USD/GBP & Individual & 4.5253 & \textbf{0.0109} \\
Gold flow & Individual & 24.0171 & \textbf{\textless{}0.001} \\
3m, FX, gold & Joint block & 21.0055 & \textbf{\textless{}0.001} \\
\bottomrule
\end{tabular*}
\begin{tablenotes}[flushleft]
\footnotesize
\item[] Bold entries: $p<0.05$ (two-sided).
\end{tablenotes}
\end{threeparttable}
\end{table}

BIC selects two lags on the differenced policy system; see Table~\ref{tab:ch2_lag}. All three companion blocks individually forecast $\Delta$Bank Rate at the 5\% level, and their joint block is highly significant. The estimated companion roots remain near the unit circle, so orthogonalized impulse responses in Appendix~\ref{app:var_setup}, Figure~\ref{fig:var_irf}, are illustrative only.

\subsubsection{Johansen cointegration (external block)}

Johansen trace tests on the four external series in levels select one cointegrating relation at the 5\% level (Table~\ref{tab:ch2_johansen}). Table~\ref{tab:johansen_vectors} reports the normalized vector (USD/GBP scaled to one). Positive loadings on imports and negative loadings on exports and reserves are consistent with long-run comovement among external-balance indicators under fixed convertibility.

\begin{table}[!t]
\centering
\small
\caption{Johansen cointegrating vector, external block ($r=1$, USD/GBP = 1)}
\label{tab:johansen_vectors}
\begin{tabular*}{\columnwidth}{@{\extracolsep{\fill}}lr}
\toprule
Variable & Vector 1 \\
\midrule
USD/GBP exchange rate & 1.000 \\
Imports & 0.163 \\
Exports & -0.162 \\
BoE reserve stock & -0.073 \\
\bottomrule
\end{tabular*}
\end{table}

\subsubsection{VECM short-run dynamics (external block)}

At $r=1$, Table~\ref{tab:vecm_granger} reports VECM Granger tests for BoE reserve stock in levels. Trade flows and the exchange rate enter significantly at conventional levels, linking merchandise payments and sterling pressure to reserve movements in the error-correction framework.

\begin{table}[!t]
\centering
\small
\begin{threeparttable}
\caption{VECM Granger causality (external block, $r=1$): forecasting BoE reserve stock}
\label{tab:vecm_granger}
\begin{tabular*}{\columnwidth}{@{\extracolsep{\fill}}lrr}
\toprule
Cause & Stat. & $p$-value \\
\midrule
USD/GBP exchange rate & 2.2820 & \textbf{0.0442} \\
Imports & 2.8704 & \textbf{0.0137} \\
Exports & 10.4190 & \textbf{\textless{}0.001} \\
\bottomrule
\end{tabular*}
\begin{tablenotes}[flushleft]
\footnotesize
\item Four-variable external block (nonstationary in levels): USD/GBP, imports, exports, and BoE reserve stock. Johansen trace selects $r=1$ (Table~\ref{tab:ch2_johansen}). Bold entries: $p<0.05$ (two-sided).
\end{tablenotes}
\end{threeparttable}
\end{table}

\subsubsection{Toda--Yamamoto causality (policy block)}

On the policy block in levels, Toda--Yamamoto Wald tests \citep{TodaYamamoto1995} with BIC-based short-run lags and one augmentation lag show that the open-market rate Granger-causes Bank Rate at the 1\% level, while exchange-rate changes and gold flow do not individually at conventional levels in this four-variable setup; see Table~\ref{tab:ty_granger}.

\begin{table}[!t]
\centering
\small
\begin{threeparttable}
\caption{Toda--Yamamoto Granger tests, policy block ($p=2$, $d_{max}=1$)}
\label{tab:ty_granger}
\begin{tabular*}{\columnwidth}{@{\extracolsep{\fill}}lrrr}
\toprule
Cause & $p$ & Total lags & $p$-value \\
\midrule
3-month market rate & 2 & 3 & \textbf{\textless{}0.001} \\
USD/GBP exchange rate & 2 & 3 & 1.0000 \\
Gold flow & 2 & 3 & 1.0000 \\
\bottomrule
\end{tabular*}
\begin{tablenotes}[flushleft]
\footnotesize
\item[] Bold entries: $p<0.05$ (two-sided).
\end{tablenotes}
\end{threeparttable}
\end{table}

\subsubsection{Single-equation MLR versus VAR in differences}

Table~\ref{tab:mlr_vs_var} compares HAC joint block tests from the seasonal ten-predictor MLR at six lags with VAR block tests on the policy differenced system at two lags. In the saturated MLR, the three-month and exchange-rate blocks are insignificant ($p \approx 0.33$ and $0.18$) and imports are marginal ($p \approx 0.062$); in the differenced VAR, the three-month rate and exchange rate enter individually at the 5\% level. Toda--Yamamoto tests assign significance primarily to the open-market rate on levels; the external VECM links trade and the exchange rate to reserve-stock dynamics rather than to Bank Rate directly.

\begin{table}[!t]
\centering
\small
\begin{threeparttable}
\caption{Forecasting $\Delta$Bank Rate: single-equation MLR ($p=6$, HAC) vs.\ VAR in differences ($p=2$, BIC)}
\label{tab:mlr_vs_var}
\begin{tabular*}{\columnwidth}{@{\extracolsep{\fill}}lrr}
\toprule
Predictor & MLR block $p$ & VAR block $p$ \\
\midrule
3-month market rate & 0.3339 & \textbf{\textless{}0.001} \\
USD/GBP exchange rate & 0.1779 & \textbf{0.0109} \\
Imports & 0.0621 & --- \\
\bottomrule
\end{tabular*}
\begin{tablenotes}[flushleft]
\footnotesize
\item[] Bold entries: $p<0.05$ (two-sided).
\end{tablenotes}
\end{threeparttable}
\end{table}

\subsubsection{Feedback from Bank Rate to markets and external variables}

The differenced policy VAR in Table~\ref{tab:ch2_granger_var} asks which shocks help forecast $\Delta$Bank Rate. Table~\ref{tab:ch2_granger_var_feedback} reports the reverse direction at the same BIC lag order ($p=2$): whether lagged $\Delta$Bank Rate helps forecast each of the other three policy-block variables. At that lag depth, none of the reverse-direction tests reject the null of no feedback at the 5\% level.

\begin{table}[!t]
\centering
\small
\begin{threeparttable}
\caption{VAR Granger causality ($p=2$ by BIC): does $\Delta$Bank Rate forecast other variables?}
\label{tab:ch2_granger_var_feedback}
\begin{tabular*}{\columnwidth}{@{\extracolsep{\fill}}lrr}
\toprule
Variable forecast & $\chi^2$ & $p$-value \\
\midrule
3-month market rate & 1.2279 & 0.2931 \\
USD/GBP exchange rate & 0.1155 & 0.8909 \\
Gold flow & 0.0560 & 0.9456 \\
\bottomrule
\end{tabular*}
\begin{tablenotes}[flushleft]
\small
\item[] Bold entries: $p<0.05$ (two-sided).
\end{tablenotes}
\end{threeparttable}
\end{table}

The saturated MLR identifies marginal forecasting content in the full dashboard. The policy VAR recovers joint bill-market and convertibility dynamics. The external VECM links trade, sterling pressure, and reserve-stock adjustment in levels. Section~\ref{sec:spec_checks} reports checks on convertibility measurement and discrete rate increases. Section~\ref{sec:stability_oos} reports subperiod stability and pseudo-real-time forecasting.

\subsection{Specification and measurement checks}
\label{sec:spec_checks}

Table~\ref{tab:hypotheses} maps domestic money-market stress to bill-market rates and external convertibility pressure to the exchange rate, gold flows, reserves, and trade. The confirmatory MLR in Section~\ref{sec:diagnostics} includes eleven monthly seasonal dummies by default. This subsection reports robustness on external-pressure measurement, a no-seasonal variant, and discrete rate increases.

\subsubsection{Monthly seasonality (no-seasonal robustness)}

Merchandise trade and prices follow pronounced calendar seasonality at monthly frequency. Table~\ref{tab:seasonality} compares HAC block $p$-values for all ten predictors in the seasonal baseline (Section~\ref{sec:diagnostics}) against the same specification without monthly dummies. Omitting seasonals strengthens imports, exports, reserve stock, and the six-month bill rate; with seasonals included, imports are marginal at $p \approx 0.062$, the six-month rate is marginal at $p \approx 0.052$, and reserve stock is insignificant at $L=6$. Exchange-rate changes remain insignificant in both specifications.

\begin{table}[!t]
\centering
\small
\begin{threeparttable}
\caption{HAC block $p$-values: canonical MLR with monthly seasonal dummies vs.\ no-seasonal robustness ($p=6$)}
\label{tab:seasonality}
\begin{tabular*}{\columnwidth}{@{\extracolsep{\fill}}lrr}
\toprule
Predictor & With season & No season \\
\midrule
BoE Reserve Notes Coin diff & 0.3314 & \textbf{0.0407} \\
USD/GBP exchange rate & 0.1779 & 0.3168 \\
Exports & \textbf{0.0059} & \textbf{\textless{}0.001} \\
Gilt price index & 0.2968 & 0.1361 \\
Gold flow & \textbf{\textless{}0.001} & \textbf{\textless{}0.001} \\
Imports & 0.0621 & \textbf{0.0026} \\
Industrial share prices & \textbf{0.0016} & \textbf{\textless{}0.001} \\
3-month market rate & 0.3339 & 0.0849 \\
6-month market rate & 0.0517 & \textbf{0.0119} \\
Wheat price & 0.1424 & 0.1802 \\
\bottomrule
\end{tabular*}
\begin{tablenotes}[flushleft]
\footnotesize
\item[] Bold entries: $p<0.05$ (two-sided).
\end{tablenotes}
\end{threeparttable}
\end{table}

\subsubsection{Exchange-rate pressure and gold points}

Historians and target-zone studies often describe external pressure in terms of parity and gold-point bands rather than month-to-month changes in the quoted \$/\pounds{} rate \citep{Eichengreen1992,BordoMacDonald2005,Officer1996}. We construct parity deviation and pressure outside fixed $\pm 1\%$ gold points from the monthly cable rate and mint parity of \$4.866 per pound; formulae are in Appendix~\ref{app:gold_points}. Table~\ref{tab:parity_fx} compares HAC block tests when monthly changes in the quoted rate, absolute deviation from parity, or gold-point pressure enter the full predictor set. Raw monthly exchange-rate changes and absolute deviation from parity are insignificant at conventional levels in the saturated MLR; pressure outside the gold points is significant at the 5\% level. The differenced policy VAR in Section~\ref{sec:var} nevertheless assigns individual forecasting content to exchange-rate innovations when estimated jointly with bill-market rates and gold flow.

\begin{table}[!t]
\centering
\small
\begin{threeparttable}
\caption{HAC block $p$-values for exchange-rate measures ($p=6$, full predictor set)}
\label{tab:parity_fx}
\begin{tabular*}{\columnwidth}{@{\extracolsep{\fill}}lr}
\toprule
FX measure & Block $p$ (HAC) \\
\midrule
$\Delta$\$/\pounds{} (monthly) & 0.1779 \\
Outside gold points & \textbf{0.0451} \\
Abs.\ deviation from \$4.866/\pounds{} & 0.4515 \\
\bottomrule
\end{tabular*}
\begin{tablenotes}[flushleft]
\footnotesize
\item[] Bold entries: $p<0.05$ (two-sided).
\end{tablenotes}
\end{threeparttable}
\end{table}

\subsubsection{London--New York interest differential}

Under gold-standard arbitrage, external pressure may also appear in the London--New York interest spread \citep{BordoMacDonald2005,Eichengreen1992}. Table~\ref{tab:intdiff} in Appendix~\ref{app:robustness_extra} compares block tests when exchange-rate changes or the differenced London minus New York spread replaces the baseline exchange-rate term.

\subsubsection{Discrete rate increases: probit and logit}

Bank Rate often moved in discrete steps while the MLR models continuous monthly changes. Table~\ref{tab:probit_main} reports joint Wald tests on six lags in probit and logit models for $\mathbb{P}(\Delta\text{Bank Rate}>0)$, using the full ten-predictor set as regressors but without monthly seasonal dummies. None of the three predictors highlighted in that table---the three-month bill rate, the cable exchange rate, and imports---pass at the 5\% level for rate increases in this discrete-outcome specification, which differs from the linear $\Delta$Bank Rate blocks in Section~\ref{sec:diagnostics}. Full specification output appears in Appendix~\ref{app:robustness_extra}, Table~\ref{tab:probit}.

\begin{table}[!t]
\centering
\small
\begin{threeparttable}
\caption{Probit and logit: joint significance of six lags for Bank Rate increases}
\label{tab:probit_main}
\begin{tabular*}{\columnwidth}{@{\extracolsep{\fill}}lrr}
\toprule
Predictor & Probit $p$ & Logit $p$ \\
\midrule
3-month market rate & 0.1867 & 0.1651 \\
USD/GBP exchange rate & 0.1153 & 0.1027 \\
Imports & 0.9782 & 0.9635 \\
\bottomrule
\end{tabular*}
\begin{tablenotes}[flushleft]
\footnotesize
\item Outcome: $\mathbb{P}(\Delta\text{Bank Rate}>0)$; 521 observations, 146 increases. Full specification in Table~\ref{tab:probit}.
\item[] Bold entries: $p<0.05$ (two-sided).
\end{tablenotes}
\end{threeparttable}
\end{table}

\subsection{Temporal stability and out-of-sample performance}
\label{sec:stability_oos}

Section~\ref{sec:results} shows that no predictor is significant in every rolling window and that different subperiods activate different variables. This subsection reports rolling designs with fewer parameters, formal break tests, and pseudo-real-time forecasts at alternative lag lengths.

\subsubsection{Rolling windows with fewer parameters}

The main rolling screen (Table~\ref{tab:rolling_window_freq}) uses three lags, ten predictors, and 37 seven-year windows on the Jan 1870--Dec 1913 panel. Table~\ref{tab:rolling_lighter} in Appendix~\ref{app:robustness_extra} repeats the exercise with lighter specifications using two or three lags and six core predictors. At $p=3$, the three-month rate is significant in 5 of 37 windows (14\%) in both the ten-predictor baseline and the six-predictor core design.

\subsubsection{Structural breaks}
\label{sec:structural_breaks}

Rolling windows describe gradual drift. Table~\ref{tab:structural_breaks} in Appendix~\ref{app:robustness_extra} reports a Bai--Perron-style search \citep{BaiPerron2003} and a Quandt--Andrews supremum Chow test \citep{QuandtAndrews1993,Chow1960}. BIC selects no interior breaks. The supremum Chow statistic peaks in the late 1870s. Chow tests at the Panic of 1873, the Baring Crisis of 1890, and the Panic of 1907 do not reject coefficient stability at the 5\% level where the test is defined. Coefficient variation across subperiods appears gradual rather than concentrated in discrete crisis regimes for the core MLR.

\subsubsection{Out-of-sample forecasting}
\label{sec:oos}

Granger and block tests are in-sample. We use an expanding-window pseudo experiment, re-estimating each month with past data only and forecasting $\Delta$Bank Rate one month ahead. We compare an autoregressive (AR) benchmark with core and full MLRs at maximum lags of one, four, and six months.

Table~\ref{tab:oos_by_lag} reports the same lag depths as Table~\ref{tab:lag_sensitivity}. The AR benchmark attains the lowest root mean squared error (RMSE) at six lags; the full ten-predictor MLR performs modestly worse at that depth and the six-predictor core MLR trails both. At one lag, information criteria favor parsimony in sample as well. Multivariate models raise the area under the receiver operating characteristic curve for classifying rate increases at shorter lag depths in some specifications; full metrics are in Appendix~\ref{app:robustness_extra}.

\begin{table}[!t]
\centering
\small
\caption{Out-of-sample RMSE by maximum lag (BIC favors $p=1$; confirmatory MLR uses $p=6$)}
\label{tab:oos_by_lag}
\begin{tabular*}{\columnwidth}{@{\extracolsep{\fill}}lrrr}
\toprule
Model & $p=1$ & $p=4$ & $p=6$ \\
\midrule
AR on Delta Bank Rate & 0.4846 & 0.4764 & 0.4736 \\
Core MLR (six predictors) & 0.5020 & 0.6554 & 0.6332 \\
Full MLR (ten predictors) & 0.6186 & 0.7357 & 0.7735 \\
\bottomrule
\end{tabular*}
\end{table}

\subsection{Time Series Visualizations}

Appendix~\ref{app:timeseries} presents dual-axis plots for all seven predictors with at least one Bonferroni- or FDR-significant exploratory hit (exports, industrial share prices, gold flow, BoE reserve stock, the three- and six-month bill rates, and wheat). Each figure shows six lags from one through six months, with the lagged change in the predictor on the left axis and $\Delta$Bank Rate on the right; two predictors appear per appendix page. The panels are descriptive complements to the regressions and help assess whether exploratory leads correspond to visible comovement rather than isolated outliers.

Episodes of tighter bill-market rates often precede or coincide with positive $\Delta$Bank Rate months around the Baring Crisis of 1890 and the run-up to 1907 \citep{Flandreau1995,Ugolini2016}, although the seasonal confirmatory MLR assigns only marginal joint content to bill-market rates once both maturities enter together; see Table~\ref{tab:bill_rate_robustness} and Figures~\ref{fig:ts_market3m} and~\ref{fig:ts_market6m}. Export and gold-flow movements line up with rate increases in several stress intervals (Figures~\ref{fig:ts_exports} and~\ref{fig:ts_goldflow}), while raw cable exchange-rate changes never pass Bonferroni or FDR screens and are omitted here. Formal break tests in Section~\ref{sec:stability_oos} do not identify discrete coefficient shifts at the pre-specified crisis dates.

\subsection{Summary of Key Findings}
\label{sec:summary_findings}

Table~\ref{tab:summary} summarizes exploratory and rolling-screen significance against confirmatory block tests. The seasonal confirmatory MLR in Section~\ref{sec:diagnostics} highlights exports, gold flow, and industrial share prices. Imports and the six-month bill rate are marginal. Raw exchange-rate changes do not pass joint blocks. Gold-point and London--New York spread measures are significant in Section~\ref{sec:spec_checks}. Section~\ref{sec:var} reports the domestic--external core in differenced, levels-based, and feedback representations.

These headline block results are supported by the robustness checks reported above. The export and gold-flow conclusions hold at maximum lags one, four, and six because joint HAC block tests pass at all three depths (Table~\ref{tab:lag_sensitivity}). Exports, gold flow, and industrial share prices also pass in the no-seasonal MLR (Table~\ref{tab:seasonality}), so those blocks do not depend on seasonal adjustment. Each London bill maturity is highly significant when entered alone (Table~\ref{tab:bill_rate_robustness}). Gold-point pressure passes joint tests when it replaces the raw cable rate (Table~\ref{tab:parity_fx}), and gold flow and the exchange rate forecast $\Delta$Bank Rate in the policy VAR (Table~\ref{tab:mlr_vs_var}).

\begin{table}[!t]
\centering
\small
\begin{threeparttable}
\caption{Summary of Granger causality results (Jan 1870--Dec 1913)}
\label{tab:summary}
\setlength{\tabcolsep}{3pt}
\begin{tabular*}{\columnwidth}{@{\extracolsep{\fill}}@{}lcccc@{}}
\toprule
Predictor & Bonf. & FDR & Roll. & Block \\
\midrule
3-month market rate & No & Some & 5/37 & No \\
6-month market rate & No & Yes & 8/37 & Marg. \\
Gold flow & No & Yes & 7/37 & Yes \\
BoE reserve stock & Yes & Yes & 13/37 & No \\
Exports & Yes & Yes & 8/37 & Yes \\
Industrial shares & Yes & Yes & 7/37 & Yes \\
Imports & No & No & 7/37 & Marg. \\
Gilt prices & No & No & 11/37 & No \\
USD/GBP (cable) & No & No & 6/37 & No \\
Wheat price & No & Some & 5/37 & No \\
\bottomrule
\end{tabular*}
\begin{tablenotes}[flushleft]
\footnotesize
\item Bonf.: Bonferroni; FDR: false discovery rate; exploratory min-$p$ screen, $L=1,\ldots,6$.
\item Roll.: uncorrected windows (Table~\ref{tab:rolling_window_freq}).
\item Block: confirmatory HAC joint test at $L=6$ (Section~\ref{sec:diagnostics}); Yes = significant at 5\%.
\item Gold-point pressure: Table~\ref{tab:parity_fx}. No-seasonal robustness: Table~\ref{tab:seasonality}. Probit for rate increases: Table~\ref{tab:probit_main}.
\end{tablenotes}
\end{threeparttable}
\end{table}

\section{Discussion}
\label{sec:discussion}

This section synthesizes the patterns in Section~\ref{sec:results}---which monthly indicators forecast changes in Bank Rate---and asks what they imply for the historiographical debate in Section~\ref{sec:background}. Section~\ref{sec:discussion_discretion} reads the estimates against sharpened mechanical and ad hoc benchmarks and interprets the middle pattern as systematic discretion. Section~\ref{sec:discussion_dashboard} interprets the seasonal confirmatory MLR as the Court's multivariate information dashboard. Section~\ref{sec:discussion_channels} places domestic money-market and external convertibility signals in that framework across subperiods and stress episodes.

\subsection{Systematic discretion: the headline answer}
\label{sec:discussion_discretion}

A mechanical reading of classical gold-standard central banking---associated with \citet{Bloomfield1959} and \citet{Nurkse1944}---implies a narrow, relatively rigid reaction function: Bank Rate adjusts predictably in response to aggregate gold flows and reserve positions. \citet{Sayers1976} and \citet{Flandreau1995}, by contrast, stress discretion: judgment over domestic financial conditions and crisis management as motivations for rate changes. On the first view, monthly data should show a few dominant predictors in nearly every subperiod with fixed relative weights; on the second, unstructured discretion would leave little recurrent structure across rolling subsamples. The estimates fit neither extreme.

The rolling-window screen is the clearest test of stability over time. Across 37 seven-year subsamples, no predictor is significant in every window, even without multiple-testing correction; see Table~\ref{tab:rolling_window_freq}. BoE reserve stock registers in only 13 of 37 uncorrected windows---the highest share---while gold flow appears in 7 and the three-month bill rate in 5. Under Bonferroni correction, the strongest result is 3 of 37 windows for gilt prices. That dispersion is difficult to reconcile with a single frozen channel operating unchanged from 1870 to 1913.

The seasonal confirmatory baseline points the same way. HAC joint block tests at six lags assign significance to exports, gold flow, and industrial share prices; imports at $p \approx 0.062$ and the six-month bill rate at $p \approx 0.052$ are marginal; BoE reserve stock, the three-month rate, gilt prices, and raw cable exchange-rate changes do not pass at the 5\% level; see Table~\ref{tab:summary} and Section~\ref{sec:diagnostics}. A story in which reserve movements alone governed Bank Rate would understate the joint content of trade and asset-price indicators in the saturated reaction function.

The evidence does not support unstructured discretion either. Exports and gold flow are jointly significant at maximum lags one, four, and six in the seasonal confirmatory specification; industrial share prices enter from four lags onward; see Table~\ref{tab:lag_sensitivity}. That recurrence across lag depths is difficult to reconcile with purely idiosyncratic discretion. Rolling subsamples show a wider repertoire: reserves and gilt prices appear most often without correction, while trade, gold, bill-market rates, and industrial share prices each register in a substantial minority of windows; see Table~\ref{tab:rolling_window_freq}.

Formal break tests on the core MLR are consistent with gradual re-weighting rather than discrete regime shifts at the Baring Crisis of 1890 or the Panic of 1907: Bai--Perron model selection with BIC chooses zero interior breaks, the Quandt--Andrews supremum Chow statistic peaks in the late 1870s rather than at those crisis dates, and Chow tests at 1890 and 1907 do not reject coefficient stability at conventional levels; see Table~\ref{tab:structural_breaks} in Appendix~\ref{app:robustness_extra}. That pattern complements the rolling screen; the main evidence for recurrent structure remains the confirmatory lag-sensitivity results and rolling frequencies above.

Taken together, the pattern fits systematic discretion: the Directorate appears to have drawn repeatedly on a menu of domestic and external indicators---trade, gold, reserve and asset-price positions, bill-market rates, and episodically convertibility measures---while the relative prominence of each line shifted across subperiods and model representations. No single predictor dominates every era under strict correction, yet several recur often enough to rule out a purely ad hoc account. The following subsections unpack how domestic money-market and external convertibility channels enter that menu.

\subsection{The Court's information dashboard}
\label{sec:discussion_dashboard}

This subsection addresses the institutional question: when the Court reviewed reserves, the exchanges, and the discount market on Thursdays, which monthly ledger lines would have carried marginal advance information about the next published change in Bank Rate once the rest of the dashboard is held constant? The seasonal confirmatory MLR in Section~\ref{sec:diagnostics} is our reconstruction of that question, asking which observable indicators still forecast $\Delta$Bank Rate when all ten predictors enter together \citep{Ugolini2016,Accominotti2021,Granger1969}.

The significant and marginal blocks in Table~\ref{tab:summary} appear in an interpretable manner. Exports are significant while imports are only marginal, consistent with payment outflows and merchandise settlement mattering more than import volumes alone in the joint information set \citep{Accominotti2021}. Gold flow enters significantly even though reserve stock does not, separating convertibility settlement through the bullion account from Banking Department reserve position as a marginal monthly lead for $\Delta$Bank Rate. Industrial share prices pass while gilt and wheat proxies do not, pointing to City financial expectations rather than a generic asset or commodity index. Reserve stock and bill-market rates remain institutionally central---the Court plainly watched both---but in the saturated six-lag specification their incremental forecasting content is weaker or obscured than trade, gold flow, and industrial prices imply. Scaled HAC coefficients for the block-significant predictors are modest month-to-month---roughly three to seven basis points on $\Delta$Bank Rate per one within-sample standard deviation of the predictor; see Table~\ref{tab:effect_sizes}---which is appropriate for a rate that moved in discrete steps and often remained unchanged; the point is recurrent informational content across decades.

On the three-month and six-month bill rates, each London maturity is highly significant when entered independently, although neither is clearly significant when both enter the baseline MLR; see Table~\ref{tab:bill_rate_robustness}. This pattern fits near-collinearity between maturities; see Table~\ref{tab:ch2_vif}. On convertibility, raw month-to-month cable changes are weak in the saturated MLR, yet pressure outside the gold points is significant when substituted into the same specification; see Table~\ref{tab:parity_fx}. Exchange-rate innovations also forecast $\Delta$Bank Rate in the four-variable policy VAR; see Table~\ref{tab:mlr_vs_var}. In other words, although exchange-rate pressure appears to lack significance when summarized in the MLR alongside nine other predictors, it shows up when measured as proximity to the gold points or estimated jointly with bill-market rates and gold flows \citep{Officer1996,Eichengreen1992,BordoMacDonald2005}.

The bill and foreign exchange patterns above are the reason the policy VAR and external VECM in Section~\ref{sec:var} belong in the same discussion. The differenced policy VAR identifies short-run trends: when bill-market rates, the exchange rate, and gold flow are modeled together with Bank Rate at monthly frequency, innovations in all three forecast subsequent $\Delta$Bank Rate; see Table~\ref{tab:mlr_vs_var}. That result recovers money-market and convertibility content compressed in the saturated MLR. By contrast, the external VECM models a long-run trend: trade, sterling pressure, and reserve stock comove in levels and predict reserve-stock adjustment without re-estimating the Bank Rate reaction function. Short-run policy leads and long-run external balance therefore sit in the same institutional environment rather than contradicting one another. Together, these models show systematic discretion: the Court drew on a recurring multivariate information set---trade settlement, gold movements, City asset prices, episodic convertibility pressure, and discount-market conditions---but the relative prominence of each line depended on how tightly the regression conditioned on the rest of the ledger.

\subsection{Domestic City, external tether, and crisis episodes}
\label{sec:discussion_channels}

Finally, we address the debate of whether the Bank prioritized domestic money-market stability or external convertibility when it moved Bank Rate. A domestic reading \citep{Bagehot1873,Goodhart1972} would stress London bill rates, Banking Department reserves, and City financial conditions, while an external reading \citep{Officer1996,Eichengreen1992} would stress gold flows, sterling pressure, and merchandise payments. Our monthly estimates support neither as an exclusive priority frozen from 1870 to 1913.

The two channels instead appear with state-dependent weights that differ by representation and subperiod. In the saturated seasonal MLR, trade settlement, gold movements, and industrial share prices carry the clearest joint blocks, while bill-market rates and reserve stock are institutionally central yet weaker or obscured once all ten predictors enter together; see Section~\ref{sec:discussion_dashboard}. In rolling seven-year windows, reserve stock and gilt prices register most often without correction, while trade, gold flow, bill-market rates, and industrial share prices each recur in a substantial minority of subsamples; see Table~\ref{tab:rolling_window_freq}. The differenced policy VAR recovers bill-market and exchange-rate content in a compact short-run system, and Toda--Yamamoto tests on the policy block still assign significance to the open-market rate on levels; see Table~\ref{tab:ty_granger}. The external VECM links trade and sterling pressure to reserve-stock adjustment over the long run. From an aggregate summary, we can thus see that domestic and external indicators sat on the same menu, but their relative prominence shifted with model conditioning and subperiod.

Stress episodes illustrate this shifting emphasis across the sample. Around the Baring Crisis of 1890, tighter bill-market conditions often preceded or coincided with positive $\Delta$Bank Rate months, consistent with City-centered discretionary management rather than a mechanical gold-outflow rule alone \citep{Flandreau1995,Sayers1976}. In the run-up to 1907, external gold drains and domestic tight money moved together, as when American banks drew gold from London and the Bank raised Bank Rate to 7\% \citep{Ugolini2016}. Figures~\ref{fig:ts_exports} and~\ref{fig:ts_goldflow} show that export and gold-flow movements comove more visibly with official rate changes in some crisis intervals than across the full sample. Formal break tests nevertheless do not identify sharp coefficient shifts at 1890 or 1907 in the core MLR; see Table~\ref{tab:structural_breaks} in Appendix~\ref{app:robustness_extra}. Crises appear to have intensified signals the Court already watched rather than replacing one channel with another.

Taken together, the pattern is closer to \citet{Flandreau1995} and \citet{Ugolini2016} than to a single-priority story. The Bank drew on domestic and external indicators when adjusting policy, with emphasis that varied across tranquil and stressed intervals and across model representations, consistent with the systematic discretion summarized in Section~\ref{sec:discussion_discretion}.

\subsection{Limitations}

\begin{itemize}
    \item Forecasting versus structure: Granger and related tests establish temporal precedence and predictive content, not the Bank's motives or a structural policy rule. Significant variables may proxy for unobserved stress such as gold flows or bill-market illiquidity \citep{Ugolini2016,Accominotti2021}. Bank Rate was unchanged in about two-fifths of months; Table~\ref{tab:probit_main} supplements the main $\Delta$Bank Rate models with probit results for rate increases.

    \item Monthly frequency: The Court set Bank Rate on a weekly rhythm, but monthly data are the feasible limit for this multivariate panel. Intra-month dynamics are not captured.

    \item Multiple testing and specification search: Bonferroni and FDR adjust across ten predictors within a given lag depth, but not across six lag specifications, composite heatmap minima, or 37 rolling windows. Seven-year rolling windows with many regressors also leave limited degrees of freedom. Confirmatory block tests, multivariate comparisons in Section~\ref{sec:var}, and Table~\ref{tab:mlr_vs_var} are reported alongside exploratory significance counts.

    \item Lag depth and out-of-sample performance: Joint block tests at six lags follow the pre-specified exploratory maximum; see Table~\ref{tab:lag_sensitivity}. Information criteria favor shorter orders, and an AR benchmark attains the lowest out-of-sample RMSE at comparable depths; see Table~\ref{tab:oos_by_lag}. Exports and gold flow are jointly significant at one, four, and six lags in the seasonal baseline; imports are marginal only at six lags. Table~\ref{tab:seasonality} reports no-seasonal robustness. The contribution centers on which indicators carried informative leads for $\Delta$Bank Rate.

    \item Gold-point bands: External-pressure robustness uses fixed $\pm 1\%$ bands around mint parity; see Appendix~\ref{app:gold_points}, rather than time-varying shipping and insurance costs as in \citet{Officer1996}.

\end{itemize}

\section{Conclusion}

The monthly evidence assembled here is most consistent with systematic discretion as the best summary of Bank Rate policy from 1870 to 1913. The Court returned repeatedly to a familiar multivariate information set---trade settlement, gold movements, City asset prices, and episodic bill-market and convertibility pressures---whose relative emphasis shifted across subperiods rather than following a single frozen channel or unstructured case-by-case judgment.

More generally, the London panel offers one concrete picture of discretionary central banking under convertibility: official rate changes tracked several observable public market lines jointly, with weights that varied as money-market and exchange conditions changed. The same joint monthly forecasting design could be applied to other historical central banks---the Banque de France, the Reichsbank, and comparable core issuers in the European comparative literature \citep{Morys2013}---once monthly trade, reserve, bill-rate, and exchange series are assembled for each institution. Those exercises would test whether the middle pattern documented here---recurring multivariate leads with state-dependent weights rather than a single frozen channel or unstructured ad hoc policy---characterized gold-standard rate-setting more broadly, or whether it reflects London's particular role as the era's principal international financial center. Either outcome would sharpen the rules-versus-discretion debate beyond a single national case; the comparative evidence is left to future work.

\section*{Acknowledgements}

The author thanks Dr.\ Shihao Yang (H.~Milton Stewart School of Industrial and Systems Engineering, Georgia Institute of Technology) for foundational training, methodology, and logic developed during prior collaborative work, which informed the computational framework used in this study. The author also thanks Professor Svetlozar Rachev (Department of Mathematics and Statistics, Texas Tech University) for substantive comments on early drafts and for guidance on paper structure, the time-series literature, and applied econometric practice. All errors and interpretations remain solely the author's own.

\section*{Declaration of generative AI and AI-assisted technologies in the manuscript preparation process}

During the preparation of this work the author used Cursor IDE to help polish the manuscript and to help develop and edit replication and build scripts under the author's direction. After using this tool, the author reviewed and edited the content as needed and takes full responsibility for the content of the published article.

\balance

\section*{Data availability}
The monthly panel series, differenced estimation samples, and replication code needed to reproduce the results are available at Mendeley Data: \url{https://data.mendeley.com/preview/ccy5z9nmjt?a=65db9fd9-b076-4223-827d-83b495df7356}. Variable construction and public supplementary sources are documented in Appendix~\ref{app:data_sources}.

\appendix
\onecolumn
\raggedbottom
{\setlength{\intextsep}{6pt plus 2pt minus 2pt}
\setlength{\textfloatsep}{6pt plus 2pt minus 2pt}
\setlength{\floatsep}{6pt plus 2pt minus 2pt}
\setlength{\abovecaptionskip}{4pt}
\setlength{\belowcaptionskip}{2pt}
\titlespacing*{\section}{0pt}{10pt plus 0pt}{4pt plus 0pt}
\titlespacing*{\subsection}{0pt}{8pt plus 0pt}{4pt plus 0pt}
\section{Data sources and series mapping}
\label{app:data_sources}

\par\vspace{2pt}\noindent\begin{minipage}{\linewidth}
\centering
Table~\ref{tab:nber_mapping} summarizes the monthly series in the Jan 1870--Dec 1913 panel. Granger regressions at six lags employ 521 observations after differencing and lag truncation.

\footnotesize
\setlength{\tabcolsep}{4pt}
\captionof{table}{Series mapping for the Jan 1870--Dec 1913 panel}
\label{tab:nber_mapping}
\begin{tabular}{@{}>{\raggedright\arraybackslash}p{0.30\linewidth}>{\raggedright\arraybackslash}p{0.22\linewidth}>{\raggedright\arraybackslash}p{0.42\linewidth}@{}}
\toprule
Series & Source ID & Description \\
\midrule
Bank Rate & NBER m13013 & BoE minimum official discount rate (\% p.a.) \\
3-month market rate & NBER m13016 & London open-market discount rate (\% p.a.) \\
6-month market rate & Millennium M9 col.\ 8 & Prime six-month bill rate (Nishimura 1971) \\
USD/GBP exchange rate & FRED USUKFXUKM & Cable rate, dollars per pound sterling (monthly) \\
Gold flow & NBER m14108 & Net gold bullion and specie trade (\pounds{} thousands) \\
BoE reserve stock & BoE weekly BD col.\ 5 & Banking Department reserve stock (month-end, \pounds{}m) \\
Imports & NBER m07029 & UK imports, value (\pounds{} millions) \\
Exports & NBER m07024 & UK exports, value (\pounds{} millions) \\
Wheat price & NBER m04002 & Wheat prices (shillings per quarter) \\
Industrial share prices & NBER m11012a & Industrial ordinary share price index \\
Gilt price index & NBER m11019 & London fixed-interest (gilt) security prices \\
\bottomrule
\end{tabular}

\end{minipage}\par\vspace{4pt}

\section{Exploratory Granger detail}
\label{app:exploratory}

Tables~\ref{tab:exploratory_lag_detail} and \ref{tab:exploratory_comprehensive} complete the full-sample enumeration at maximum lags $L=1,\ldots,6$ summarized in Section~\ref{sec:results}. Exports, gold flow, industrial share prices, and BoE reserve stock accumulate the most Bonferroni and FDR hits; the six-month bill rate registers more exploratory hits than the three-month rate.

\par\vspace{2pt}\noindent\begin{minipage}{\linewidth}
\centering
\scriptsize
\captionof{table}{Exploratory full-sample Granger screen: significant predictor--lag pairs by maximum lag depth ($L$)}
\label{tab:exploratory_lag_detail}
\begin{tabular*}{\linewidth}{@{\extracolsep{\fill}}clclr}
\toprule
Max lag $L$ & Predictor & Lag & $p$-value & Level \\
\midrule
4 & 3-month market rate & 4 & 0.0284 & FDR \\
5 & 3-month market rate & 4 & 0.0379 & Uncorrected \\
1 & 6-month market rate & 1 & 0.0144 & FDR \\
2 & 6-month market rate & 1 & 0.0180 & Uncorrected \\
3 & 6-month market rate & 1 & 0.0368 & Uncorrected \\
4 & 6-month market rate & 1 & 0.0242 & FDR \\
5 & 6-month market rate & 1 & 0.0178 & FDR \\
5 & 6-month market rate & 5 & 0.0223 & FDR \\
6 & 6-month market rate & 1 & 0.0336 & Uncorrected \\
6 & 6-month market rate & 5 & 0.0280 & Uncorrected \\
3 & BoE reserve stock & 3 & 0.0356 & Uncorrected \\
4 & BoE reserve stock & 4 & 0.0003 & Bonferroni \\
5 & BoE reserve stock & 4 & 0.0041 & Bonferroni \\
6 & BoE reserve stock & 4 & 0.0072 & FDR \\
1 & Exports & 1 & 0.0005 & Bonferroni \\
2 & Exports & 1 & 0.0005 & Bonferroni \\
3 & Exports & 1 & 0.0003 & Bonferroni \\
4 & Exports & 1 & 0.0084 & FDR \\
5 & Exports & 1 & 0.0008 & Bonferroni \\
5 & Exports & 5 & 0.0012 & Bonferroni \\
6 & Exports & 1 & 0.0010 & Bonferroni \\
6 & Exports & 5 & 0.0012 & Bonferroni \\
1 & Gold flow & 1 & 0.0143 & FDR \\
5 & Gold flow & 1 & 0.0128 & FDR \\
5 & Gold flow & 5 & 0.0104 & FDR \\
6 & Gold flow & 1 & 0.0194 & FDR \\
6 & Gold flow & 5 & 0.0171 & FDR \\
4 & Industrial share prices & 4 & 0.0164 & FDR \\
5 & Industrial share prices & 4 & 0.0073 & FDR \\
6 & Industrial share prices & 3 & 0.0387 & FDR \\
6 & Industrial share prices & 4 & 0.0016 & Bonferroni \\
3 & USD/GBP exchange rate & 3 & 0.0237 & Uncorrected \\
5 & USD/GBP exchange rate & 3 & 0.0436 & Uncorrected \\
4 & Wheat price & 3 & 0.0108 & FDR \\
\bottomrule
\multicolumn{5}{l}{\footnotesize Each row is a term--lag pair significant at the stated level in the MLR with all ten predictors and lags $1,\ldots,L$.} \\
\multicolumn{5}{l}{\footnotesize Bonferroni and FDR corrections are applied to the predictor-level screening summaries at each lag depth.} \\
\end{tabular*}
\end{minipage}\par\vspace{4pt}

\par\vspace{2pt}\noindent\begin{minipage}{\linewidth}
\centering
\small
\captionof{table}{Exploratory screen: count of significant term--lag instances by correction (across max lags $L=1,\ldots,6$)}
\label{tab:exploratory_comprehensive}
\begin{tabular*}{\linewidth}{@{\extracolsep{\fill}}lccc}
\toprule
Predictor & Bonferroni & FDR & Uncorrected \\
\midrule
3-month market rate & 0 & 1 & 1 \\
6-month market rate & 0 & 4 & 4 \\
USD/GBP exchange rate & 0 & 0 & 2 \\
Gold flow & 0 & 5 & 0 \\
BoE reserve stock & 2 & 1 & 1 \\
Imports & 0 & 0 & 0 \\
Exports & 7 & 1 & 0 \\
Wheat price & 0 & 1 & 0 \\
Gilt price index & 0 & 0 & 0 \\
Industrial share prices & 1 & 3 & 0 \\
\bottomrule
\multicolumn{4}{l}{\footnotesize Counts sum term--lag hits across maximum-lag specifications $L=1,\ldots,6$.} \\
\end{tabular*}
\end{minipage}\par\vspace{4pt}

\section{Confirmatory regression detail}
\label{app:confirmatory}

Table~\ref{tab:ch2_canonical_hac} reports Newey--West HAC coefficients for the open-market rate, the USD/GBP exchange rate, and imports in the canonical ten-predictor MLR ($L=6$, $N=521$). Table~\ref{tab:ch2_lag_ic} lists information criteria by maximum lag length. Tables~\ref{tab:ch2_ljungbox} and \ref{tab:ch2_vif} report residual autocorrelation and collinearity diagnostics. Bill-rate robustness and effect-size tables appear in Section~\ref{sec:diagnostics}.

\par\vspace{2pt}\noindent\begin{minipage}{\linewidth}
\centering
\small
\captionof{table}{Canonical Granger regression ($p=6$) with monthly seasonal dummies: coefficients on selected predictors of $\Delta$Bank Rate (Newey--West HAC, 6 lags)}
\label{tab:ch2_canonical_hac}
\begin{tabular*}{\linewidth}{@{\extracolsep{\fill}}llrrrr}
\toprule
Predictor & Lag & Coef. & HAC SE & $p$ (HAC) & Block $p$ (HAC) \\
\midrule
3-month market rate & 1 & -0.0924 & 0.1300 & 0.4776 & 0.3339 \\
3-month market rate & 2 & -0.0829 & 0.1158 & 0.4745 &  \\
3-month market rate & 3 & -0.1409 & 0.1175 & 0.2311 &  \\
3-month market rate & 4 & 0.1006 & 0.1007 & 0.3183 &  \\
3-month market rate & 5 & 0.0920 & 0.1344 & 0.4939 &  \\
3-month market rate & 6 & -0.1265 & 0.1021 & 0.2160 &  \\
\addlinespace
USD/GBP exchange rate & 1 & 0.9672 & 1.1152 & 0.3863 & 0.1779 \\
USD/GBP exchange rate & 2 & -0.0798 & 0.7984 & 0.9204 &  \\
USD/GBP exchange rate & 3 & 1.5881 & 0.9419 & 0.0925* &  \\
USD/GBP exchange rate & 4 & 1.0258 & 0.6143 & 0.0957* &  \\
USD/GBP exchange rate & 5 & -1.0247 & 0.6774 & 0.1310 &  \\
USD/GBP exchange rate & 6 & -0.4747 & 1.3413 & 0.7236 &  \\
\addlinespace
Imports & 1 & 0.0001 & 0.0004 & 0.8064 & 0.0621 \\
Imports & 2 & -0.0012 & 0.0007 & 0.0905* &  \\
Imports & 3 & -0.0009 & 0.0006 & 0.1574 &  \\
Imports & 4 & -0.0000 & 0.0005 & 0.9977 &  \\
Imports & 5 & -0.0001 & 0.0006 & 0.8954 &  \\
Imports & 6 & -0.0010 & 0.0004 & 0.0069*** &  \\
\addlinespace
\bottomrule
\multicolumn{6}{l}{\footnotesize Stars on $p$ (HAC) per lag: $^{***}p<0.01$, $^{**}p<0.05$, $^{*}p<0.10$. Block $p$ tests all six lags jointly.} \\
\end{tabular*}
\end{minipage}\par\vspace{4pt}

\par\vspace{2pt}\noindent\begin{minipage}{\linewidth}
\centering
\small
\captionof{table}{Information criteria by maximum lag $p$ (canonical MLR, all ten predictors)}
\label{tab:ch2_lag_ic}
\begin{tabular*}{\linewidth}{@{\extracolsep{\fill}}lrrrrr}
\toprule
$p$ & $N$ & AIC & BIC & HQIC & Adj.\ $R^2$ \\
\midrule
1$^\ddagger$ & 526 & 830.4 & 928.5 & 868.8 & 0.267 \\
2 & 525 & 835.7 & 980.6 & 892.4 & 0.273 \\
3 & 524 & 833.9 & 1025.7 & 909.0 & 0.287 \\
4$^\dagger$ & 523 & 814.4 & 1053.0 & 907.9 & 0.316 \\
5 & 522 & 822.7 & 1108.0 & 934.4 & 0.317 \\
6 & 521 & 819.3 & 1151.3 & 949.3 & 0.321 \\
\bottomrule
\multicolumn{6}{l}{\footnotesize $^\dagger$ Lowest AIC ($p=4$); $^\ddagger$ lowest BIC ($p=1$); lowest HQIC ($p=1$). Main specification uses $p=6$ ($N=521$).} \\
\end{tabular*}
\end{minipage}\par\vspace{4pt}

\par\vspace{2pt}\noindent\begin{minipage}{\linewidth}
\centering
\small
\captionof{table}{Ljung--Box tests on ordinary least squares (OLS) residuals, canonical MLR ($p=6$)}
\label{tab:ch2_ljungbox}
\begin{tabular*}{\linewidth}{@{\extracolsep{\fill}}lrr}
\toprule
Null lags & $Q$ & $p$-value \\
\midrule
6 & 1.85 & 0.9331 \\
12 & 6.04 & 0.9138 \\
24 & 17.32 & 0.8347 \\
\bottomrule
\end{tabular*}
\end{minipage}\par\vspace{4pt}

\par\vspace{2pt}\noindent\begin{minipage}{\linewidth}
\centering
\small
\captionof{table}{Maximum variance inflation factor (VIF) by predictor (across its six lags), canonical MLR ($p=6$)}
\label{tab:ch2_vif}
\begin{tabular*}{\linewidth}{@{\extracolsep{\fill}}lr}
\toprule
Variable & Max VIF \\
\midrule
3-month market rate & 10.1 \\
6-month market rate & 9.5 \\
Gold flow & 1.9 \\
BoE reserve stock & 1.8 \\
Exports & 1.6 \\
Gilt price index & 1.4 \\
Industrial share prices & 1.4 \\
Wheat price & 1.3 \\
\bottomrule
\multicolumn{2}{l}{\footnotesize VIF $>10$ suggests severe collinearity among lag columns.} \\
\end{tabular*}
\end{minipage}\par\vspace{4pt}

\section{VAR setup and lag selection}
\label{app:var_setup}

Table~\ref{tab:ch2_adf} reports integration orders for the policy VAR block; Table~\ref{tab:ch2_adf_vecm} and Table~\ref{tab:ch2_johansen} cover the external VECM block. Table~\ref{tab:ch2_lag} reports lag-order selection on the differenced policy system. Main-text VECM, Toda--Yamamoto, and cointegrating-vector tables appear in Section~\ref{sec:var}; feedback Granger tests appear in Table~\ref{tab:ch2_granger_var_feedback}.

\par\vspace{2pt}\noindent\begin{minipage}{\linewidth}
\centering
\small
\captionof{table}{Unit-root tests: policy VAR block (1870--1913)}
\label{tab:ch2_adf}
\setlength{\tabcolsep}{3pt}
\begin{tabular*}{\linewidth}{@{\extracolsep{\fill}}lrrl}
\toprule
Variable & ADF stat. & ADF $p$ & Order \\
\midrule
Bank Rate & -4.3813 & 0.0003 & I(0) \\
3-month market rate & -3.7275 & 0.0037 & I(0) \\
USD/GBP exchange rate & -2.4214 & 0.1358 & I(1) \\
Gold flow & -4.6563 & 0.0001 & I(0) \\
\bottomrule
\end{tabular*}
\end{minipage}\par\vspace{4pt}

\par\vspace{2pt}\noindent\begin{minipage}{\linewidth}
\centering
\small
\captionof{table}{Unit-root tests: external VECM block (1870--1913)}
\label{tab:ch2_adf_vecm}
\setlength{\tabcolsep}{3pt}
\begin{tabular*}{\linewidth}{@{\extracolsep{\fill}}lrrl}
\toprule
Variable & ADF stat. & ADF $p$ & Order \\
\midrule
USD/GBP exchange rate & -2.4214 & 0.1358 & I(1) \\
Imports & -1.7575 & 0.4017 & I(1) \\
Exports & 0.6009 & 0.9876 & I(1) \\
BoE reserve stock & -1.6279 & 0.4686 & I(1) \\
\bottomrule
\end{tabular*}
\end{minipage}\par\vspace{4pt}

\par\vspace{2pt}\noindent\begin{minipage}{\linewidth}
\centering
\small
\begin{threeparttable}
\caption{Johansen trace test: external-adjustment block (levels, 1870--1913)}
\label{tab:ch2_johansen}
\setlength{\tabcolsep}{3pt}
\begin{tabular*}{\linewidth}{@{\extracolsep{\fill}}lrr}
\toprule
$r$ & Trace stat. & 95\% crit. \\
\midrule
0 & \textbf{122.325} & 47.854 \\
1 & 20.442 & 29.796 \\
2 & 8.297 & 15.494 \\
3 & 0.024 & 3.842 \\
\bottomrule
\end{tabular*}
\begin{tablenotes}[flushleft]
\footnotesize
\item All four series are nonstationary in levels by ADF (Table~\ref{tab:ch2_adf_vecm}). Trace tests select $r=1$ at the 5\% level. Bold trace statistics reject $H_0$: rank $\leq r$ at 5\%.
\end{tablenotes}
\end{threeparttable}
\end{minipage}\par\vspace{4pt}

\par\vspace{2pt}\noindent\begin{minipage}{\linewidth}
\centering
\small
\captionof{table}{VAR lag-order selection on first differences (policy block)}
\label{tab:ch2_lag}
\setlength{\tabcolsep}{3pt}
\begin{tabular*}{\linewidth}{@{\extracolsep{\fill}}lr}
\toprule
criterion & optimal lag \\
\midrule
AIC & 12 \\
BIC & 2 \\
FPE & 12 \\
HQIC & 11 \\
\bottomrule
\end{tabular*}
\end{minipage}\par\vspace{4pt}

\par\vspace{4pt}\noindent\begin{minipage}{\linewidth}
\centering
\includegraphics[width=0.85\textwidth]{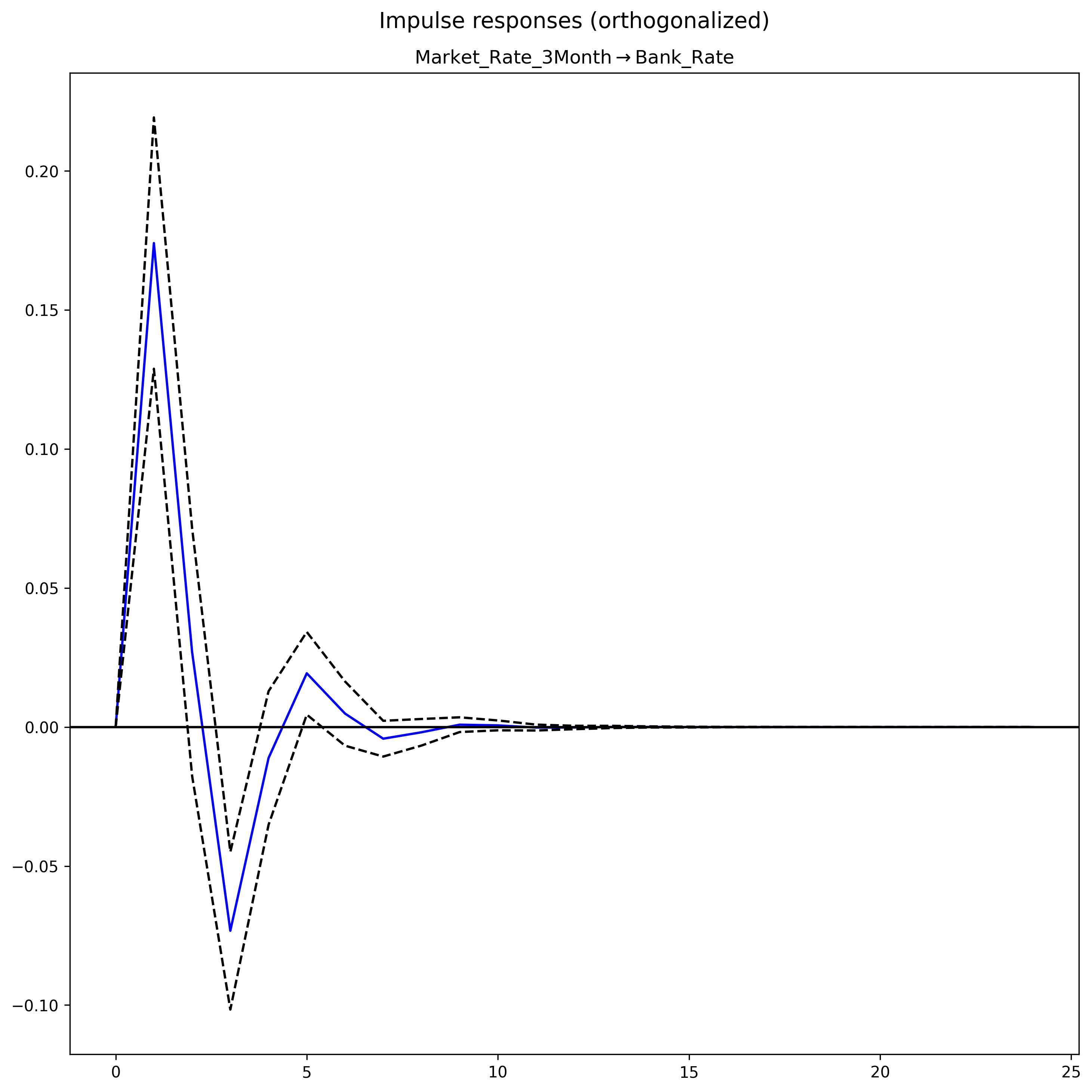}
\captionof{figure}{Orthogonalized impulse response of $\Delta$Bank Rate to a shock in $\Delta$(three-month market rate), BIC VAR with two lags (illustrative; companion roots near the unit circle).}
\label{fig:var_irf}
\end{minipage}\par\vspace{6pt}

\section{Supplementary robustness}
\label{app:robustness_extra}

Tables~\ref{tab:parity_fx}--\ref{tab:chow_crisis} and \ref{tab:oos_auc_by_lag}--\ref{tab:oos_forecast} complement the robustness checks in Sections~\ref{sec:spec_checks} and~\ref{sec:stability_oos}. Main-text tables report the core VAR/VECM comparisons, seasonality, and out-of-sample RMSE by lag length (Table~\ref{tab:oos_by_lag}); Tables~\ref{tab:oos_auc_by_lag} and~\ref{tab:oos_forecast} report area under the curve (AUC), mean absolute error (MAE), and related forecast metrics.

Table~\ref{tab:reserve_cover_proxy} compares HAC block tests for BoE reserve stock and a reserves-to-trade ratio constructed from the same series and monthly trade flows. Both measures yield similar joint significance at the 5\% level in this sample.

\subsection{Gold-point pressure construction}
\label{app:gold_points}

Let $s_t$ denote the monthly cable sterling--dollar rate in dollars per pound (FRED \texttt{USUKFXUKM}, aligned with NBER \texttt{m14105}). Mint parity is $\bar{s} = 4.866$ USD per pound. A fixed gold-point band of $\pm 1\%$ around parity defines import and export gold points $g^{\mathrm{imp}}_t = \bar{s}(1 + 0.01)$ and $g^{\mathrm{exp}}_t = \bar{s}(1 - 0.01)$, following the stylized target-zone bandwidth in \citet{BordoMacDonald2005}. Deviation from parity is $d_t = s_t - \bar{s}$; pressure outside the band is $p_t = \max(|d_t| - 0.01\bar{s},\, 0)$. First differences $\Delta d_t$ and $\Delta p_t$ enter robustness block tests in Table~\ref{tab:parity_fx}. This construction does not vary shipping, insurance, or handling costs over time as in \citet{Officer1996}; it is a transparent monthly proxy rather than a full Officer-style time-varying band.

\par\vspace{2pt}\noindent\begin{minipage}{\linewidth}
\centering
\small
\begin{threeparttable}
\caption{HAC block $p$-values: reserve levels vs.\ reserves/trade proxy ($L=6$)}
\label{tab:reserve_cover_proxy}
\begin{tabular*}{\linewidth}{@{\extracolsep{\fill}}lr}
\toprule
Reserve measure & Block $p$ (HAC) \\
\midrule
BoE reserve stock (first difference) & 0.3314 \\
Reserves / (imports + exports), first difference & 0.3314 \\
\bottomrule
\end{tabular*}
\begin{tablenotes}[flushleft]
\small
\item[] Bold entries: $p<0.05$ (two-sided).
\end{tablenotes}
\end{threeparttable}
\end{minipage}\par\vspace{4pt}

\par\vspace{2pt}\noindent\begin{minipage}{\linewidth}
\centering
\small
\begin{threeparttable}
\caption{HAC block $p$-values: $\Delta\$/\pounds{}$ vs.\ London--NY 3m spread ($p=6$)}
\label{tab:intdiff}
\begin{tabular*}{\linewidth}{@{\extracolsep{\fill}}lr}
\toprule
External measure & Block $p$ (HAC) \\
\midrule
$\Delta$\$/\pounds{} (monthly) & 0.1779 \\
$\Delta$(London 3m $-$ NY call) & \textbf{0.0033} \\
\bottomrule
\end{tabular*}
\begin{tablenotes}[flushleft]
\footnotesize
\item[] Bold entries: $p<0.05$ (two-sided).
\end{tablenotes}
\end{threeparttable}
\end{minipage}\par\vspace{4pt}

\par\vspace{2pt}\noindent\begin{minipage}{\linewidth}
\centering
\small
\captionof{table}{Probit and logit: joint significance of six lags ($\mathbb{P}(\Delta\text{Bank Rate}>0)$)}
\label{tab:probit}
\begin{tabular*}{\linewidth}{@{\extracolsep{\fill}}llrr}
\toprule
Outcome & Predictor & Probit $p$ & Logit $p$ \\
\midrule
Rate increase & 3-month market rate & 0.1867 & 0.1651 \\
Rate increase & USD/GBP exchange rate & 0.1153 & 0.1027 \\
Rate increase & Imports & 0.9782 & 0.9635 \\
any change & 3-month market rate & 0.4321 & 0.4355 \\
any change & USD/GBP exchange rate & 0.0939 & 0.0955 \\
any change & Imports & 0.8774 & 0.8893 \\
\bottomrule
\end{tabular*}
\end{minipage}\par\vspace{4pt}

\par\vspace{2pt}\noindent\begin{minipage}{\linewidth}
\centering
\small
\captionof{table}{Rolling windows: \% with uncorrected significance for $\Delta$(3m rate)}
\label{tab:rolling_lighter}
\begin{tabular*}{\linewidth}{@{\extracolsep{\fill}}lrrr}
\toprule
Specification & Windows & Sig.\ count & \% sig. \\
\midrule
Baseline ($p=3$, 10 predictors) & 37 & 5 & 13.5 \\
$p=2$, 10 predictors & 37 & 1 & 2.7 \\
$p=3$, 6 core predictors & 37 & 5 & 13.5 \\
$p=2$, 6 core predictors & 37 & 2 & 5.4 \\
\bottomrule
\end{tabular*}
\end{minipage}\par\vspace{4pt}

\par\vspace{2pt}\noindent\begin{minipage}{\linewidth}
\centering
\small
\captionof{table}{Structural break tests on the core MLR ($p=6$; six predictors plus $\Delta$Bank Rate lags)}
\label{tab:structural_breaks}
\begin{tabular*}{\linewidth}{@{\extracolsep{\fill}}lll}
\toprule
Method & Result & Break date \\
\midrule
Bai--Perron (BIC, 0--3 breaks) & 0 break(s) selected & --- \\
Quandt--Andrews sup-Chow (1 break) & sup-F = 2.58 & 1879-10-01 \\
\bottomrule
\end{tabular*}
\end{minipage}\par\vspace{4pt}

\par\vspace{2pt}\noindent\begin{minipage}{\linewidth}
\centering
\small
\captionof{table}{Chow tests for coefficient stability at pre-specified crisis months (core MLR, $p=6$)}
\label{tab:chow_crisis}
\begin{tabular*}{\linewidth}{@{\extracolsep{\fill}}llrrrr}
\toprule
Episode & Break date & Chow $F$ & $p$-value & $N$ pre & $N$ post \\
\midrule
Panic of 1873 & 1873-09-01 & nan & nan & 38 & 483 \\
Baring crisis 1890 & 1890-11-01 & 0.9480 & 0.5597 & 244 & 277 \\
Panic of 1907 & 1907-10-01 & 0.6120 & 0.9659 & 447 & 74 \\
\bottomrule
\end{tabular*}
\end{minipage}\par\vspace{4pt}

\par\vspace{2pt}\noindent\begin{minipage}{\linewidth}
\centering
\small
\captionof{table}{Out-of-sample area under the curve (AUC) for Bank Rate increases by maximum lag}
\label{tab:oos_auc_by_lag}
\begin{tabular*}{\linewidth}{@{\extracolsep{\fill}}lrrr}
\toprule
Model & $p=1$ & $p=4$ & $p=6$ \\
\midrule
AR on Delta Bank Rate & 0.6820 & 0.7320 & 0.7370 \\
Core MLR (six predictors) & 0.6840 & 0.6610 & 0.6740 \\
Full MLR (ten predictors) & 0.6560 & 0.6590 & 0.6580 \\
\bottomrule
\end{tabular*}
\end{minipage}\par\vspace{4pt}

\par\vspace{2pt}\noindent\begin{minipage}{\linewidth}
\centering
\small
\captionof{table}{Pseudo-real-time one-step-ahead forecasts at $p=6$ (expanding window from 120 months). Columns report root mean squared error (RMSE), mean absolute error (MAE), and area under the curve (AUC).}
\label{tab:oos_forecast}
\begin{tabular*}{\linewidth}{@{\extracolsep{\fill}}lrrrrr}
\toprule
Model & $N$ & RMSE & MAE & Sign hit (\%) & AUC (increase) \\
\midrule
AR(6) on Delta Bank Rate & 400 & 0.4736 & 0.3225 & 73.0 & 0.737 \\
Core MLR (p=6, six predictors) & 400 & 0.6332 & 0.4202 & 65.7 & 0.674 \\
Full MLR (p=6, ten predictors) & 400 & 0.7735 & 0.5107 & 63.0 & 0.658 \\
\bottomrule
\end{tabular*}
\end{minipage}\par\vspace{4pt}

\section{Rolling window heatmaps}
\label{app:rolling}

Figures~\ref{fig:rolling_raw}--\ref{fig:rolling_bonf} display rolling-window $p$-values for the same 37 seven-year windows ($L=3$, ten predictors): uncorrected, FDR-adjusted, and Bonferroni-adjusted.

\par\vspace{4pt}\noindent\begin{minipage}{\linewidth}
\centering
\includegraphics[width=\textwidth]{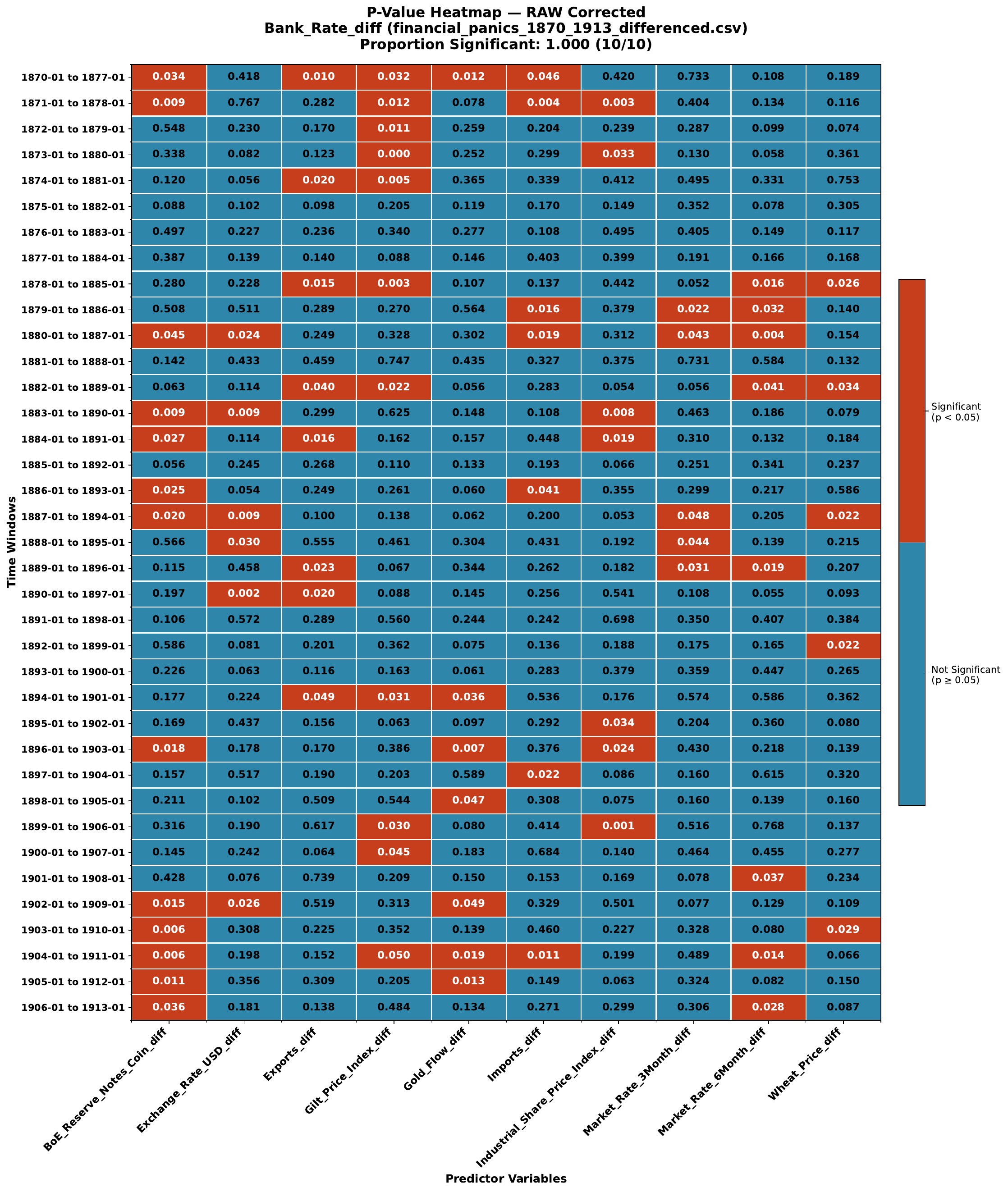}
\captionof{figure}{Rolling-window raw $p$-values (37 seven-year windows, one-year steps, $L=3$). Rows are time windows; columns are predictors. Red cells indicate $p<0.05$. \texttt{BoE\_Reserve\_Notes\_Coin\_diff} is significant in 13 of 37 windows (35.1\%).}
\label{fig:rolling_raw}
\end{minipage}\par\vspace{6pt}

\par\vspace{4pt}\noindent\begin{minipage}{\linewidth}
\centering
\includegraphics[width=\textwidth]{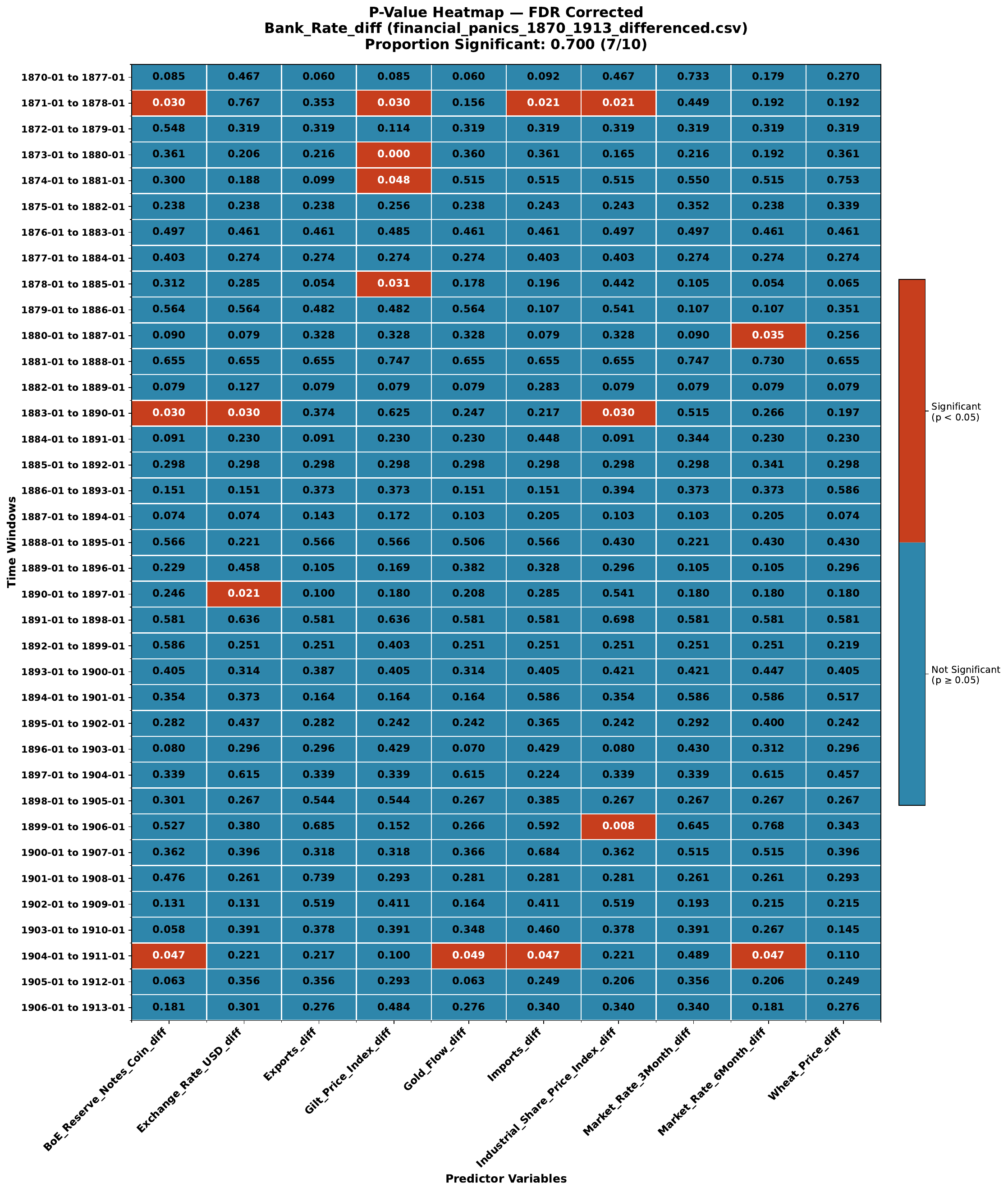}
\captionof{figure}{Rolling-window FDR-adjusted $p$-values (37 seven-year windows). \texttt{Gilt\_Price\_Index\_diff} is significant in 4 of 37 windows (10.8\%).}
\label{fig:rolling_fdr}
\end{minipage}\par\vspace{6pt}

\par\vspace{4pt}\noindent\begin{minipage}{\linewidth}
\centering
\includegraphics[width=\textwidth]{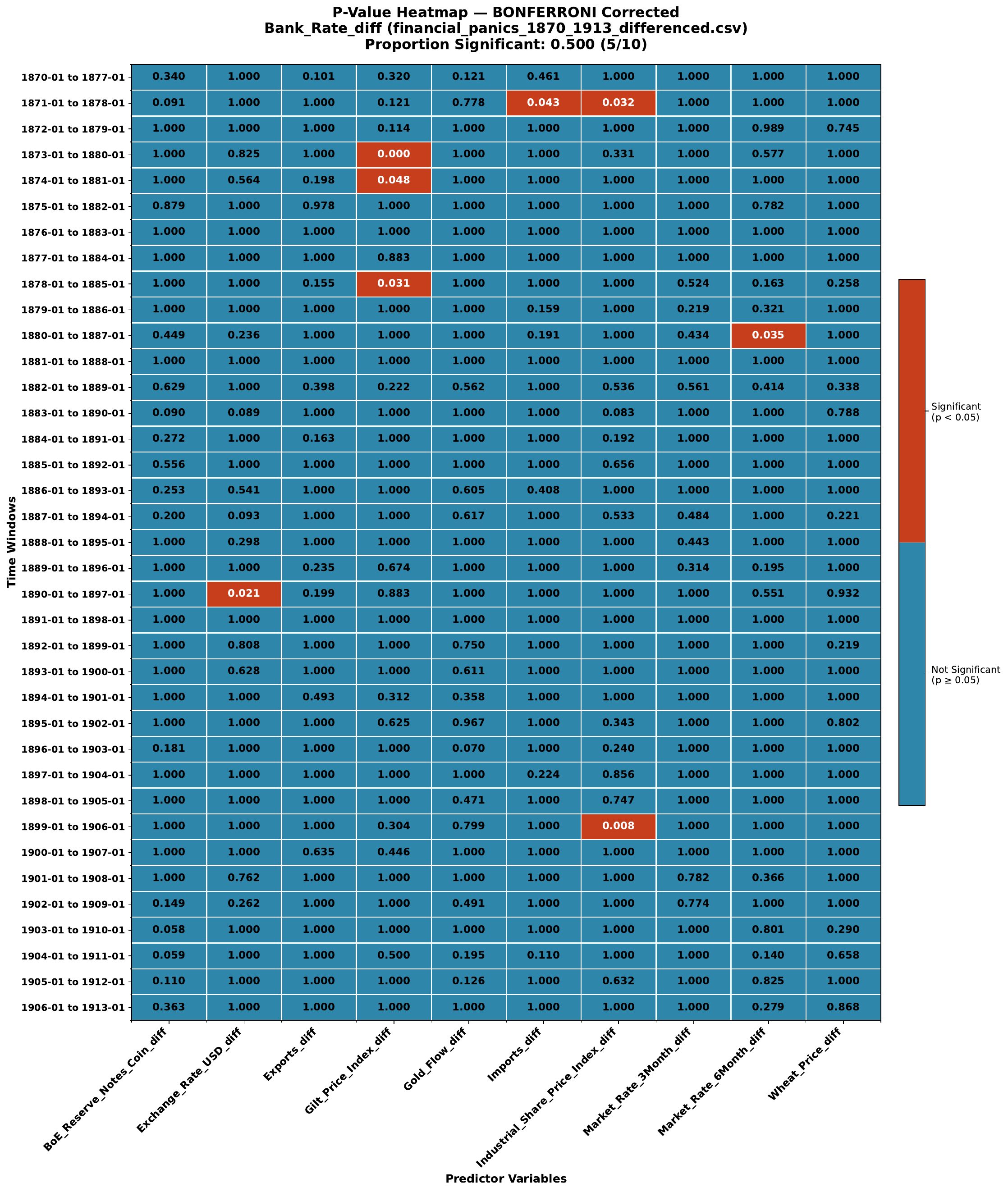}
\captionof{figure}{Rolling-window Bonferroni-adjusted $p$-values (37 seven-year windows). \texttt{Gilt\_Price\_Index\_diff} is significant in 3 of 37 windows (8.1\%).}
\label{fig:rolling_bonf}
\end{minipage}\par\vspace{6pt}

\section{Time series plots}
\label{app:timeseries}

Dual-axis plots for the seven predictors with at least one Bonferroni- or FDR-significant term--lag pair in the exploratory MLR (Tables~\ref{tab:exploratory_comprehensive} and \ref{tab:exploratory_lag_detail}). Each panel shows lags one--six; the lagged predictor is on the left axis and \texttt{Bank\_Rate\_diff} on the right. Two predictors per page.

\par\vspace{4pt}\noindent\begin{minipage}{\linewidth}
\centering
\includegraphics[width=\textwidth,height=0.36\textheight,keepaspectratio]{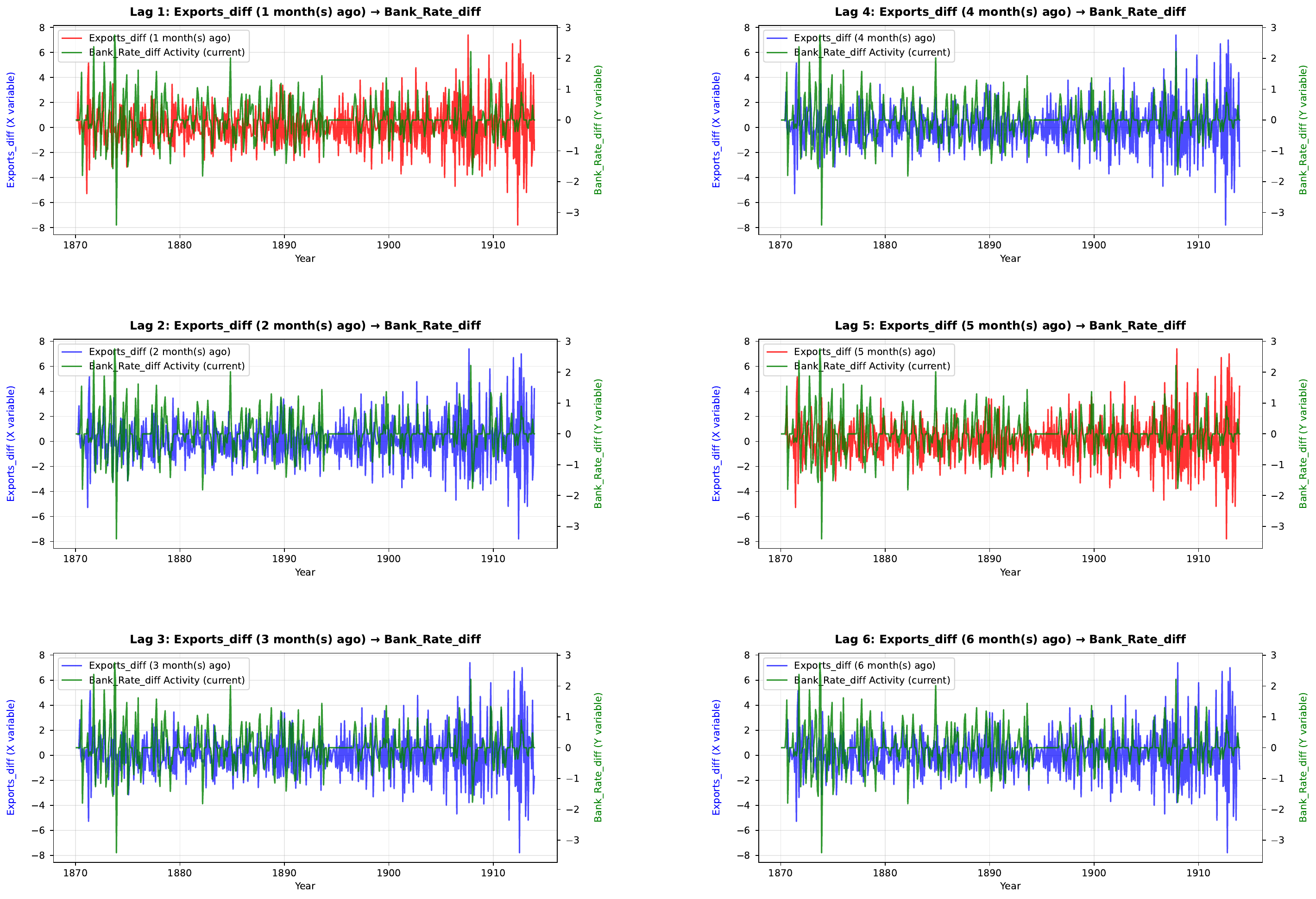}
\captionof{figure}{Exports (Bonferroni and FDR; confirmatory block significant at $L=6$).}
\label{fig:ts_exports}
\end{minipage}\par\vspace{6pt}

\par\vspace{4pt}\noindent\begin{minipage}{\linewidth}
\centering
\includegraphics[width=\textwidth,height=0.36\textheight,keepaspectratio]{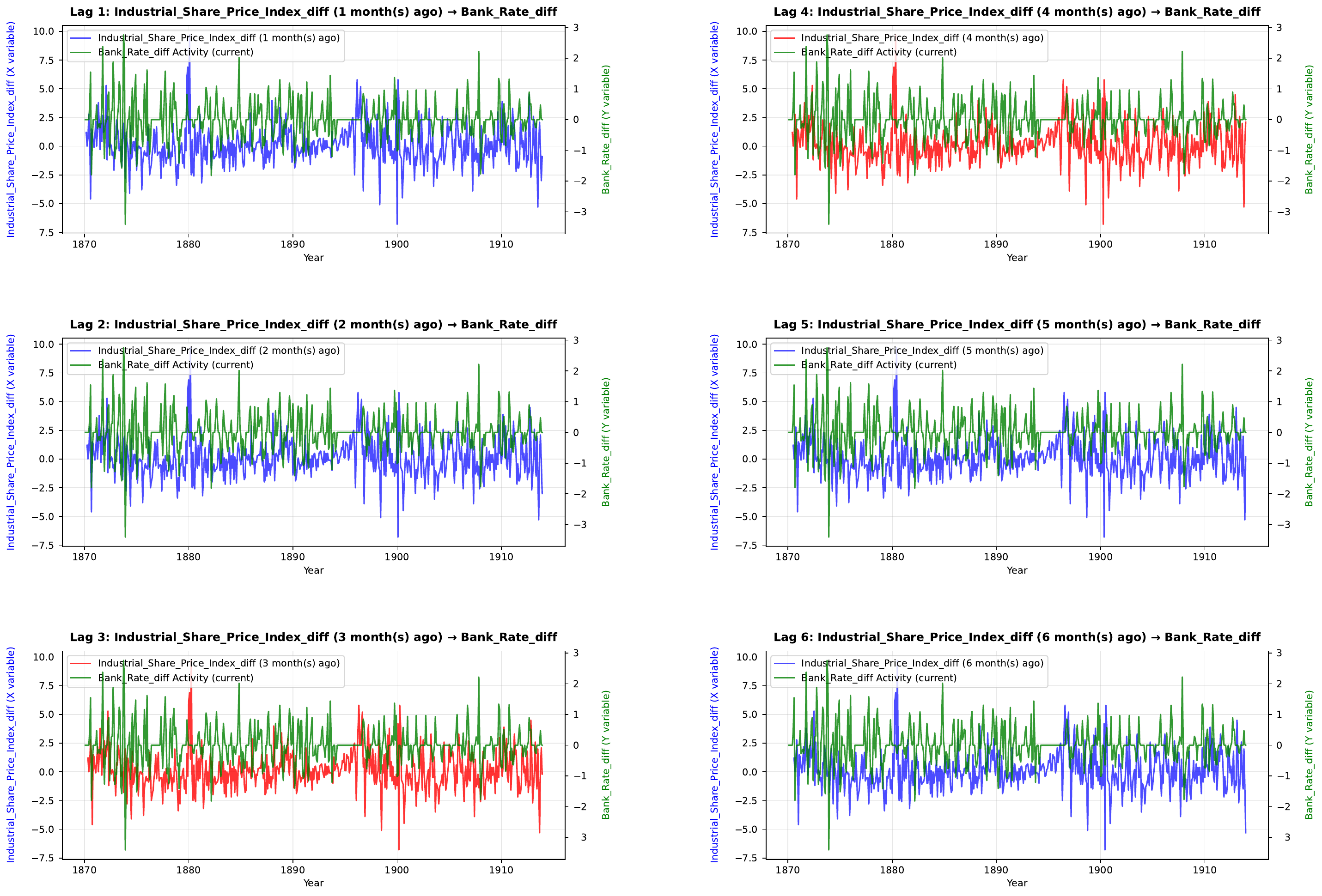}
\captionof{figure}{Industrial share prices (Bonferroni and FDR; confirmatory block significant at $L=6$).}
\label{fig:ts_industrial}
\end{minipage}\par\vspace{6pt}

\newpage

\par\vspace{4pt}\noindent\begin{minipage}{\linewidth}
\centering
\includegraphics[width=\textwidth,height=0.36\textheight,keepaspectratio]{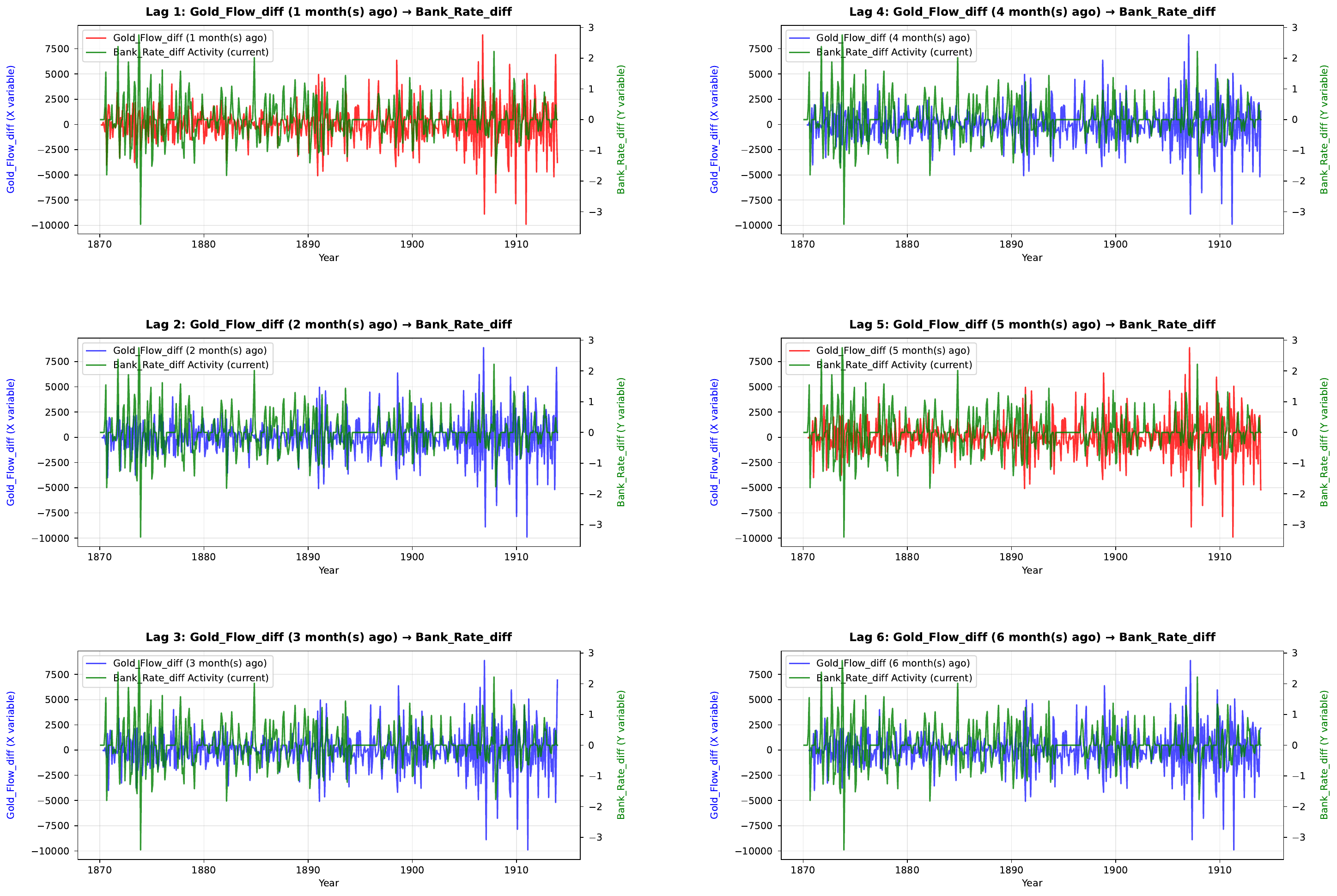}
\captionof{figure}{Gold flow (FDR; confirmatory block significant at $L=6$).}
\label{fig:ts_goldflow}
\end{minipage}\par\vspace{6pt}

\par\vspace{4pt}\noindent\begin{minipage}{\linewidth}
\centering
\includegraphics[width=\textwidth,height=0.36\textheight,keepaspectratio]{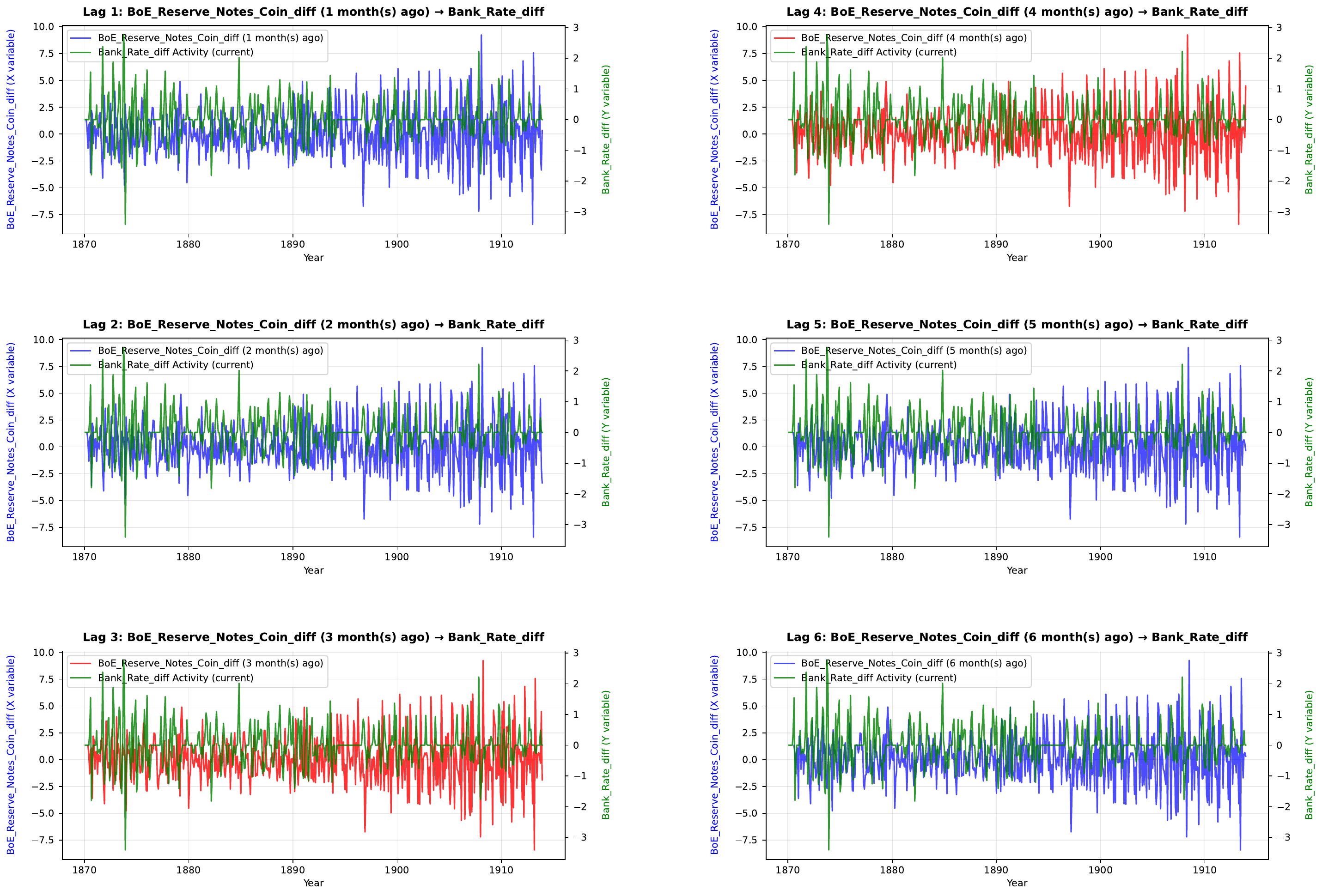}
\captionof{figure}{BoE reserve stock (Bonferroni and FDR; not confirmatory at $L=6$).}
\label{fig:ts_reserve}
\end{minipage}\par\vspace{6pt}

\newpage

\par\vspace{4pt}\noindent\begin{minipage}{\linewidth}
\centering
\includegraphics[width=\textwidth,height=0.36\textheight,keepaspectratio]{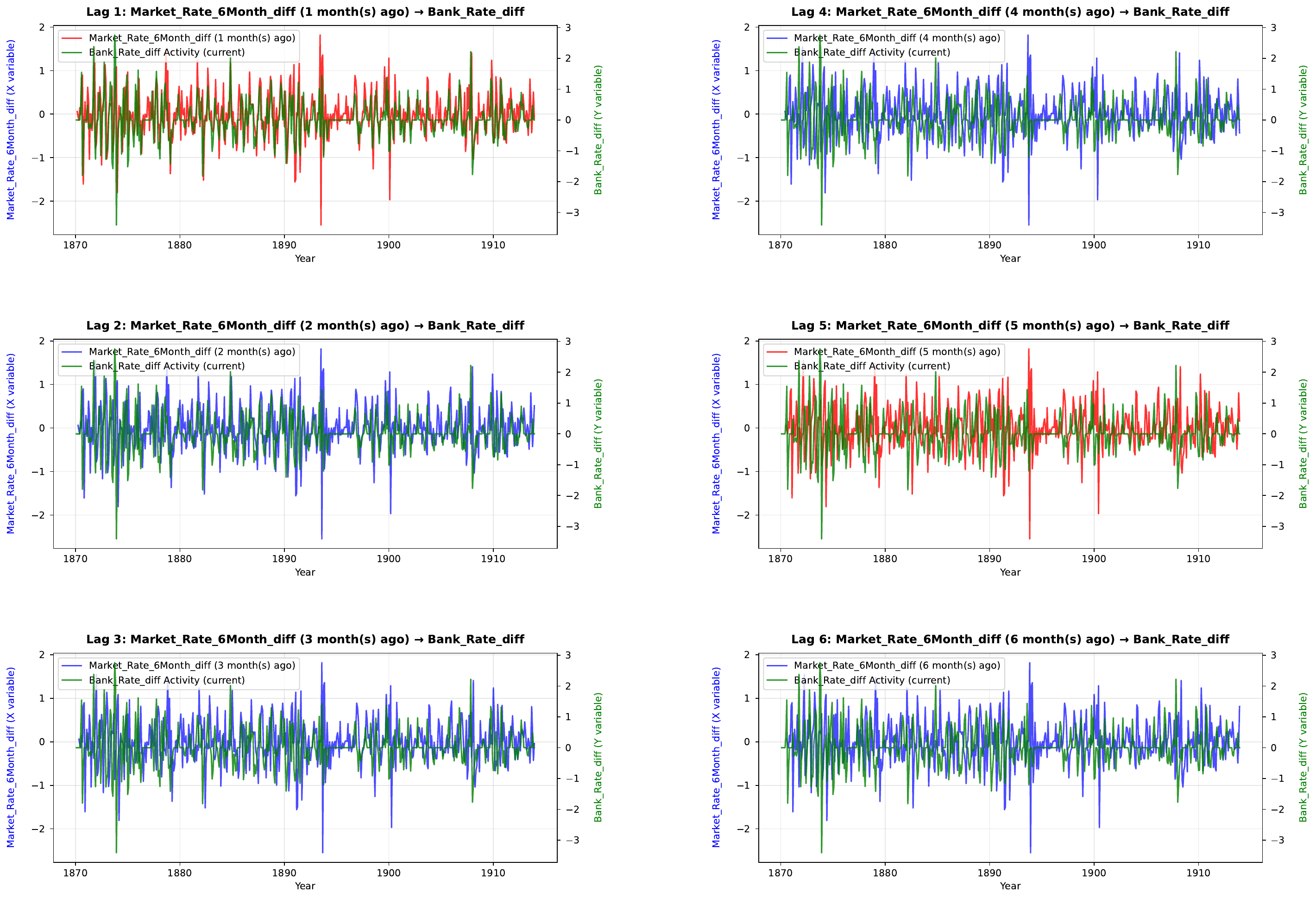}
\captionof{figure}{Six-month bill rate (FDR; marginal confirmatory block at $L=6$ when both bill rates enter).}
\label{fig:ts_market6m}
\end{minipage}\par\vspace{6pt}

\par\vspace{4pt}\noindent\begin{minipage}{\linewidth}
\centering
\includegraphics[width=\textwidth,height=0.36\textheight,keepaspectratio]{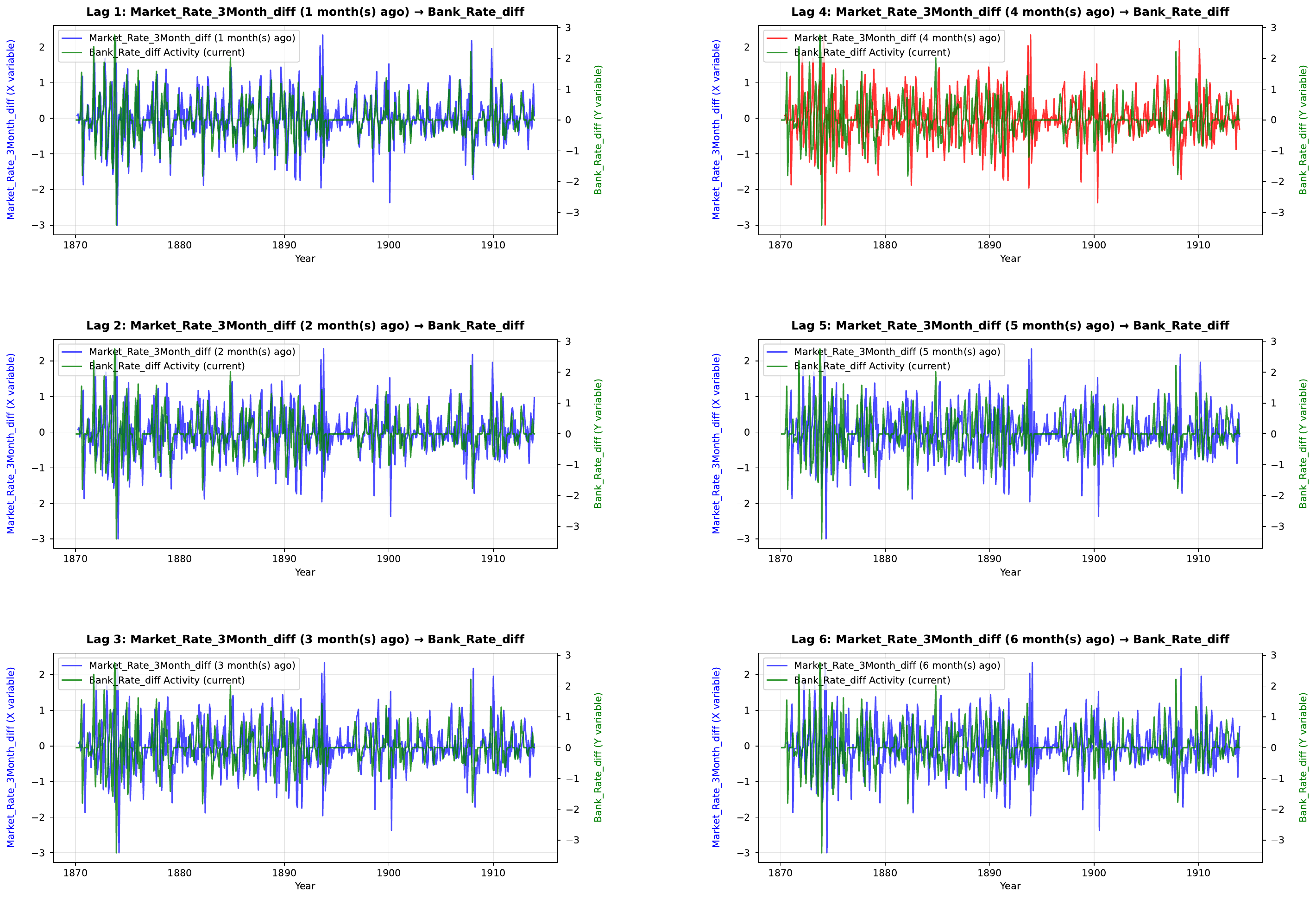}
\captionof{figure}{Three-month bill rate (FDR; split across maturities in the confirmatory MLR; Table~\ref{tab:bill_rate_robustness}).}
\label{fig:ts_market3m}
\end{minipage}\par\vspace{6pt}

\newpage

\par\vspace{4pt}\noindent\begin{minipage}{\linewidth}
\centering
\includegraphics[width=\textwidth,height=0.72\textheight,keepaspectratio]{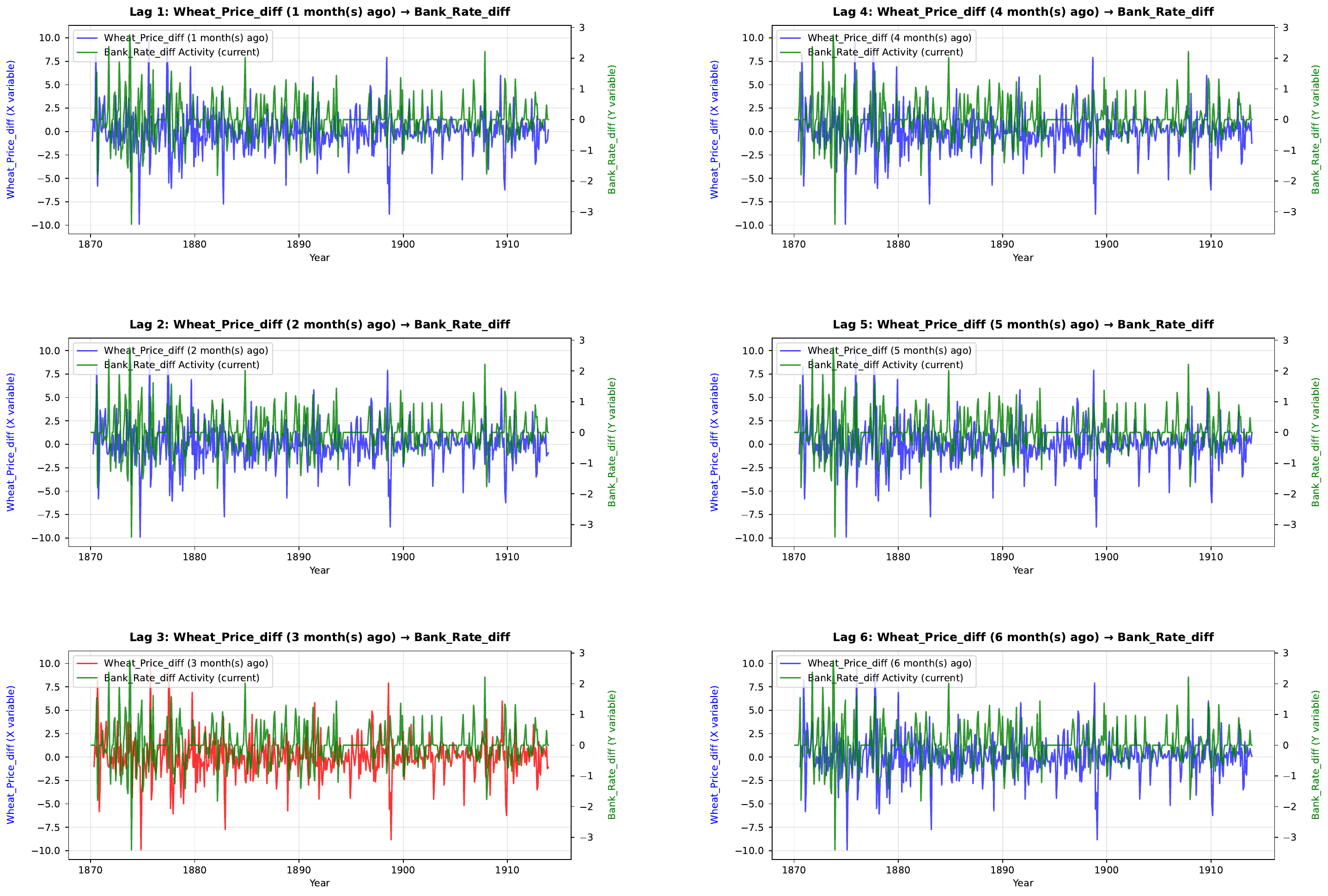}
\captionof{figure}{Wheat price (FDR only).}
\label{fig:ts_wheat}
\end{minipage}\par\vspace{6pt}

}


\begin{thebibliography}{99}

\bibitem[Bagehot(1873)]{Bagehot1873}
Bagehot, W. (1873). \textit{Lombard Street: A description of the money market}. Henry S. King \& Co. \url{https://doi.org/10.1017/cbo9781139093620}

\bibitem[Bloomfield(1959)]{Bloomfield1959}
Bloomfield, A. I. (1959). \textit{Monetary policy under the international gold standard, 1880--1914}. Federal Reserve Bank of New York.

\bibitem[Bordo \& MacDonald(2005)]{BordoMacDonald2005}
Bordo, M. D., \& MacDonald, R. (2005). Interest rate interactions in the classical gold standard, 1880--1914: Was there any monetary independence? \textit{Journal of Monetary Economics}, \textit{52}(2), 307--327. \url{https://doi.org/10.1016/j.jmoneco.2004.05.008}

\bibitem[Capie \& Webber(1985)]{CapieWebber1985}
Capie, F., \& Webber, A. (1985). \textit{A monetary history of the United Kingdom, 1870--1982}. Allen \& Unwin. \url{https://doi.org/10.4324/9781315019840}

\bibitem[Eichengreen(1992)]{Eichengreen1992}
Eichengreen, B. (1992). \textit{Golden fetters: The gold standard and the Great Depression, 1919--1939}. Oxford University Press. \url{https://doi.org/10.1093/0195101138.001.0001}

\bibitem[Eichengreen(1996)]{Eichengreen1996}
Eichengreen, B. (1996). \textit{Globalizing capital: A history of the international monetary system}. Princeton University Press. \url{https://doi.org/10.2307/j.ctt7pfmc}

\bibitem[Flandreau(1995)]{Flandreau1995}
Flandreau, M. (1995). The Bank of England and the City. In R. Roberts \& D. Kynaston (Eds.), \textit{The Bank of England: Money, power and influence 1694--1994} (pp. 152--184). Oxford University Press. \url{https://doi.org/10.1093/oso/9780198289524.003.0006}

\bibitem[Goodhart(1972)]{Goodhart1972}
Goodhart, C. A. E. (1972). \textit{The business of banking, 1891--1914}. Weidenfeld \& Nicolson.

\bibitem[Nurkse(1944)]{Nurkse1944}
Nurkse, R. (1944). \textit{International currency experience: Lessons of the inter-war period}. League of Nations.

\bibitem[Keynes(1930)]{Keynes1930}
Keynes, J. M. (1930). \textit{A treatise on money} (Vols.~1--2). Macmillan.

\bibitem[Obstfeld \& Taylor(2004)]{ObstfeldTaylor2004}
Obstfeld, M., \& Taylor, A. M. (2004). \textit{Global capital markets: Integration, crisis, and growth}. Cambridge University Press. \url{https://doi.org/10.1017/cbo9780511616525}

\bibitem[Sayers(1976)]{Sayers1976}
Sayers, R. S. (1976). \textit{The Bank of England, 1891--1944}. Cambridge University Press.

\bibitem[Accominotti et al.(2021)]{Accominotti2021}
Accominotti, O., Lucena-Piquero, D., \& Ugolini, S. (2021). The origination and distribution of money market instruments: Sterling bills of exchange during the first globalization. \textit{The Economic History Review}, \textit{74}(4), 892--921. \url{https://doi.org/10.1111/ehr.13049}

\bibitem[Lennard(2018)]{Lennard2018}
Lennard, J. (2018). Did monetary policy matter? Narrative evidence from the classical gold standard. \textit{Explorations in Economic History}, \textit{68}, 16--36. \url{https://doi.org/10.1016/j.eeh.2017.10.001}

\bibitem[Morys(2013)]{Morys2013}
Morys, M. (2013). Discount rate policy under the Classical Gold Standard: Core versus periphery (1870s--1914). \textit{Explorations in Economic History}, \textit{50}(2), 205--226. \url{https://doi.org/10.1016/j.eeh.2012.12.003}

\bibitem[Ugolini(2016)]{Ugolini2016}
Ugolini, S. (2016). Liquidity management and central bank strength: Bank of England operations reloaded, 1889--1910. \textit{Norges Bank Working Paper} 10/2016. \url{https://doi.org/10.2139/ssrn.2842571}

\bibitem[Mitchell(1988)]{Mitchell1988}
Mitchell, B. R. (1988). \textit{British historical statistics}. Cambridge University Press.

\bibitem[Granger(1969)]{Granger1969}
Granger, C. W. J. (1969). Investigating causal relations by econometric models and cross-spectral methods. \textit{Econometrica}, \textit{37}(3), 424--438. \url{https://doi.org/10.2307/1912791}

\bibitem[Granger \& Newbold(1974)]{GrangerNewbold1974}
Granger, C. W. J., \& Newbold, P. (1974). Spurious regressions in econometrics. \textit{Journal of Econometrics}, \textit{2}(2), 111--120. \url{https://doi.org/10.1016/0304-4076(74)90034-7}

\bibitem[Hamilton(1994)]{Hamilton1994}
Hamilton, J. D. (1994). \textit{Time series analysis}. Princeton University Press. \url{https://doi.org/10.1515/9780691218632}

\bibitem[Sims(1980)]{Sims1980}
Sims, C. A. (1980). Macroeconomics and reality. \textit{Econometrica}, \textit{48}(1), 1--48. \url{https://doi.org/10.2307/1912017}

\bibitem[Newey \& West(1987)]{NeweyWest1987}
Newey, W. K., \& West, K. D. (1987). A simple, positive semi-definite, heteroskedasticity and autocorrelation consistent covariance matrix. \textit{Econometrica}, \textit{55}(3), 703--708. \url{https://doi.org/10.2307/1913610}

\bibitem[Johansen(1991)]{Johansen1991}
Johansen, S. (1991). Estimation and hypothesis testing of cointegration vectors in Gaussian vector autoregressive models. \textit{Econometrica}, \textit{59}(6), 1551--1580. \url{https://doi.org/10.2307/2938278}

\bibitem[Toda \& Yamamoto(1995)]{TodaYamamoto1995}
Toda, H. Y., \& Yamamoto, T. (1995). Statistical inference in vector autoregressions with possibly integrated processes. \textit{Journal of Econometrics}, \textit{66}(1--2), 225--250. \url{https://doi.org/10.1016/0304-4076(94)01616-8}

\bibitem[Benjamini \& Hochberg(1995)]{BenjaminiHochberg1995}
Benjamini, Y., \& Hochberg, Y. (1995). Controlling the false discovery rate. \textit{Journal of the Royal Statistical Society: Series B}, \textit{57}(1), 289--300. \url{https://doi.org/10.1111/j.2517-6161.1995.tb02031.x}

\bibitem[Dickey \& Fuller(1979)]{DickeyFuller1979}
Dickey, D. A., \& Fuller, W. A. (1979). Distribution of the estimators for autoregressive time series with a unit root. \textit{Journal of the American Statistical Association}, \textit{74}(366), 427--431. \url{https://doi.org/10.1080/01621459.1979.10482531}

\bibitem[Engle \& Granger(1987)]{EngleGranger1987}
Engle, R. F., \& Granger, C. W. J. (1987). Co-integration and error correction. \textit{Econometrica}, \textit{55}(2), 251--276. \url{https://doi.org/10.2307/1913236}

\bibitem[Bai \& Perron(2003)]{BaiPerron2003}
Bai, J., \& Perron, P. (2003). Computation and analysis of multiple structural change models. \textit{Journal of Applied Econometrics}, \textit{18}(1), 1--22. \url{https://doi.org/10.1002/jae.659}

\bibitem[Quandt \& Andrews(1993)]{QuandtAndrews1993}
Quandt, R. E., \& Andrews, D. W. K. (1993). Tests for parameter instability and structural change with unknown change point. \textit{Econometrica}, \textit{61}(4), 821--856. \url{https://doi.org/10.2307/2951764}

\bibitem[Chow(1960)]{Chow1960}
Chow, G. C. (1960). Tests of equality between sets of coefficients in two linear regressions. \textit{Econometrica}, \textit{28}(3), 591--605. \url{https://doi.org/10.2307/1910133}

\bibitem[Krugman(1991)]{Krugman1991}
Krugman, P. R. (1991). Target zones and exchange rate dynamics. \textit{The Quarterly Journal of Economics}, \textit{106}(3), 669--682. \url{https://doi.org/10.2307/2937922}

\bibitem[Officer(1996)]{Officer1996}
Officer, L. H. (1996). \textit{Between the dollar-sterling gold points: Exchange rates, parity, and market behavior}. Cambridge University Press. \url{https://doi.org/10.1017/cbo9780511559723}

\bibitem[Ljung \& Box(1978)]{LjungBox1978}
Ljung, G. M., \& Box, G. E. P. (1978). On a measure of lack of fit in time series models. \textit{Biometrika}, \textit{65}(2), 297--303. \url{https://doi.org/10.1093/biomet/65.2.297}

\end{thebibliography}
\end{document}